\documentclass[sigplan,screen]{acmart}
\usepackage[ruled,lined,linesnumbered,vlined]{algorithm2e}

\SetCommentSty{mycommfont}
\DontPrintSemicolon %

\SetKwProg{Fn}{Function}{:}{}

\usepackage{amsmath}
\usepackage{amsfonts}
\usepackage{amsthm}
\usepackage{balance}
\usepackage{enumitem}
\usepackage{graphicx}
\usepackage{listings}
\usepackage{multicol}
\usepackage{multirow}
\usepackage{natbib}
\usepackage{subcaption}
\usepackage{url}
\usepackage{xspace}
\usepackage{xcolor}
\usepackage{tcolorbox}
\usepackage{colortbl}

\usepackage{tikz}
\usepackage{pgfplots}
\newcommand*\circled[1]{\tikz[baseline=(char.base)]{
    \node[shape=circle,draw,inner sep=0.5pt] (char) {\small#1};}}
\usetikzlibrary{shapes}
\usetikzlibrary{shapes.geometric}
\usetikzlibrary{arrows.meta, positioning}

\newcommand{\eg}{\mbox{e.g.}\xspace}
\newcommand{\etal}{\mbox{et al.}\xspace}

\newcommand{\ie}{\mbox{i.e.}\xspace}
\newcommand{\vs}{\mbox{vs.}\xspace}
\newcommand{\wrt}{\mbox{w.r.t.}\xspace}

\definecolor{BlindColorTolOne}{HTML}{332288}
\definecolor{BlindColorTolTwo}{HTML}{117733} %
\definecolor{BlindColorTolThree}{HTML}{44AA99}
\definecolor{BlindColorTolFour}{HTML}{88CCEE}
\definecolor{BlindColorTolFive}{HTML}{DDCC77}
\definecolor{BlindColorTolSix}{HTML}{CC6677} %
\definecolor{BlindColorTolSeven}{HTML}{AA4499}
\definecolor{BlindColorTolEight}{HTML}{882255}

\definecolor{BlindColorWongOne}{HTML}{000000} %
\definecolor{BlindColorWongTwo}{HTML}{E69F00}
\definecolor{BlindColorWongThree}{HTML}{56B4E9}
\definecolor{BlindColorWongFour}{HTML}{009E73}
\definecolor{BlindColorWongFive}{HTML}{F0E442}
\definecolor{BlindColorWongSix}{HTML}{0072B2} %
\definecolor{BlindColorWongSeven}{HTML}{D55E00}
\definecolor{BlindColorWongEight}{HTML}{CC79A7}

\definecolor{mygreen}{HTML}{02818a}

\mathchardef\mhyphen="2D

\newcounter{FindingCounter}

\newcommand{\myparagraph}[1]{
  \vspace*{0.04cm}
  \noindent \textit{\textbf{#1.}}\quad
}

\newcommand{\mycode}[1]{\texttt{#1}\xspace}

\newcommand{\projectname}[1]{\mbox{\textsf{#1}}\xspace} %
\SetKwData{AlgTrue}{true}
\SetKwData{AlgFalse}{false}

\newcommand{\proj}{\projectname{LPO}}
\newcommand{\projnofeedback}{\projectname{LPO$^-$}}

\newcommand{\llvm}{LLVM\xspace}

\newcommand{\optbenchmark}{\llvm Opt Benchmark\xspace}
\newcommand{\optbenchmarkNumProjects}{240\xspace}

\newcommand{\llms}{LLMs\xspace}
\newcommand{\llm}{LLM\xspace}
\newcommand{\opt}{\mycode{opt}}
\newcommand{\mca}{\mycode{llvm-mca}}
\newcommand{\alive}{Alive2\xspace}

\newcommand{\vgemmathree}{\hbox{gemma3:27b}\xspace}
\newcommand{\vllamathree}{\hbox{llama3.3:70b}\xspace}
\newcommand{\vgeminitwo}{\hbox{gemini-2.0-flash}\xspace}

\newcommand{\vofourmini}{\hbox{o4-mini-2025-04-16}\xspace}
\newcommand{\vgptfourone}{\hbox{gpt-4.1-2025-04-14}\xspace}

\newcommand{\gemmathree}{\hbox{Gemma3}\xspace}
\newcommand{\llamathree}{\hbox{Llama3.3}\xspace}
\newcommand{\geminitwo}{\hbox{Gemini2.0}\xspace}
\newcommand{\geminithinking}{\hbox{Gemini2.0T}\xspace}
\newcommand{\ofourmini}{\hbox{o4-mini}\xspace}
\newcommand{\gptfourone}{\hbox{GPT-4.1}\xspace}

\newcommand{\geminitwofive}{\hbox{Gemini2.5}\xspace}
\newcommand{\vgeminitwofive}{\hbox{gemini-2.5-flash-lite}\xspace}

\newcommand{\ir}{IR\xspace}
\newcommand{\souper}{Souper\xspace}
\newcommand{\minotaur}{Minotaur\xspace}
\newcommand{\rust}{Rust\xspace}
\newcommand{\clamp}{\mycode{clamp}}
\newcommand{\instcombine}{\mycode{InstCombine}}

\newcommand{\souperdefault}{Souper$_\mycode{Default}$\xspace}
\newcommand{\souperenum}{Souper$_\mycode{Enum}$\xspace}
\newcommand{\souperenumone}{Souper$_\mycode{Enum=1}$\xspace}

\newcommand{\enum}{\mycode{Enum}}
\newcommand{\default}{\mycode{Default}}

\newcommand{\numberOfExtractedSnippets}{800,000\xspace}
\newcommand{\numberOfDuplicates}{8.7 million\xspace}
\newcommand{\numberOfThroughputBenchmarks}{5,000\xspace}

\newcommand{\timeOfLPOllama}{26.2\xspace}
\newcommand{\totalTimeOfLPOllama}{36.4 hours\xspace}
\newcommand{\timeOfLPOgemini}{6.7\xspace}
\newcommand{\totalTimeOfLPOgemini}{9.3 hours\xspace}
\newcommand{\costOfLPOgemini}{5.4 USD\xspace}
\newcommand{\timeOfSouperdefault}{2.8\xspace}
\newcommand{\totalTimeOfSouperdefault}{3.9 hours\xspace}
\newcommand{\timeOfSouperenumOne}{37.2\xspace}
\newcommand{\timeOfSouperenumTwo}{144.4\xspace}
\newcommand{\timeOfSouperenumThree}{183.7\xspace}
\newcommand{\totalTimeOfSouperenumone}{52 hours\xspace}
\newcommand{\totalTimeOfSouperenumtwo}{200 hours\xspace}
\newcommand{\totalTimeOfSouperenumthree}{255 hours\xspace}

\newcommand{\numberOfSouperenumoneTimeout}{80\xspace}
\newcommand{\numberOfSouperenumtwoTimeout}{412\xspace}
\newcommand{\numberOfSouperenumthreeTimeout}{616\xspace}

\newcommand{\numberOfBenchmarks}{25\xspace}
\newcommand{\numberOfBenchmarksLLMsCanDetect}{22\xspace}

\newcommand{\numberOfBenchmarksFormalCanDetect}{15\xspace}
\newcommand{\numberOfBenchmarksMinotaurCanDetect}{3\xspace}
\newcommand{\numberOfBenchmarksFormalCannotDetect}{10\xspace}
\newcommand{\numberOfBenchmarksSouperCannotLLMsCan}{7\xspace}

\newcommand{\monthsOfExperiment}{\monthsOfExperimentNoSpace\xspace}
\newcommand{\monthsOfExperimentNoSpace}{eleven}

\newcommand{\reportedInTotal}{62\xspace}
\newcommand{\confirmedBugs}{28\xspace}
\newcommand{\fixedBugs}{13\xspace}

\newcommand{\wontFix}{3\xspace}

\newcommand{\confirmedDefaultCanFind}{3\xspace}
\newcommand{\reportedDefaultCanFind}{6\xspace}
\newcommand{\confirmedEnumCanFind}{14\xspace}
\newcommand{\reportedEnumCanFind}{20\xspace}
\newcommand{\confirmedSouperCannot}{26\xspace}
\newcommand{\confirmedMinotaurCannot}{31\xspace}
\newcommand{\confirmedMinotaurCanFind}{10\xspace}
\newcommand{\reportedMinotaurCanFind}{13\xspace}

\newcommand{\souperTimeout}{20~minutes\xspace}
\newcommand{\souperTimeoutAdj}{20-minute\xspace}

\newcommand{\souperEnumSynMaxInst}{3\xspace
}

\newcommand{\toolurl}{\url{https://github.com/uw-pluverse/lpo-artifact}\xspace}

\SetKwInOut{Input}{Input}
\SetKwInOut{Output}{Output}
\SetKw{Continue}{continue}
\SetKw{Break}{break}
\SetKwFunction{AlgAppend}{append}
\SetKwFunction{AlgAdd}{add}

\SetCommentSty{mynewcommfont}

\usepackage{pifont}
\newcommand*\colourcheck[1]{%
  \expandafter\newcommand\csname #1check\endcsname{\textcolor{#1}{\ding{51}}}%
}
\colourcheck{green}
\colourcheck{black}

\newcommand*\colourcross[1]{%
  \expandafter\newcommand\csname #1cross\endcsname{\textcolor{#1}{\ding{55}}}%
}
\colourcross{red}

\newcounter{uniqueIdCounter}

\makeatletter
\newcommand{\getOrAssignID}[1]{%
  \ifcsname myMap@#1\endcsname%
    \csname myMap@#1\endcsname%
  \else%
    \stepcounter{uniqueIdCounter}%
    \expandafter\xdef\csname myMap@#1\endcsname{\theuniqueIdCounter}%
    \theuniqueIdCounter%
  \fi%
}
\makeatother

\newcommand{\issueId}[1]{\href{https://github.com/llvm/llvm-project/issues/#1}{#1}}

\newcommand{\anonymizedIssueId}[1]{\ifx\useAnonymizedIssueId\undefined%
  \issueId{#1}%
\else%
  \getOrAssignID{#1}%
\fi}

\newcommand{\alivebuglink}{\ifx\useAnonymizedIssueId\undefined
  \url{https://github.com/AliveToolkit/alive2/issues/1229}
\else
  https://anonymous.for.review
\fi
}

\newcommand{\wontfixone}{Issue \anonymizedIssueId{130954}\xspace}
\newcommand{\wontfixtwo}{Issue \anonymizedIssueId{132628}\xspace}
\newcommand{\wontfixthree}{Issue \anonymizedIssueId{167199}\xspace}

\lstdefinelanguage{Rust}{%
  sensitive%
, morecomment=[l]{//}%
, morecomment=[s]{/*}{*/}%
, moredelim=[s][{\itshape\color[rgb]{0,0,0.75}}]{\#[}{]}%
, morestring=[b]{"}%
, alsodigit={}%
, alsoother={}%
, alsoletter={!}%
, morekeywords={break, continue, else, for, if, in, loop, match, return, while}  %
, morekeywords={as, const, let, move, mut, ref, static}  %
, morekeywords={dyn, enum, fn, impl, Self, self, struct, trait, type, union, use, where}  %
, morekeywords={crate, extern, mod, pub, super}  %
, morekeywords={unsafe}  %
, morekeywords={abstract, alignof, become, box, do, final, macro, offsetof, override, priv, proc, pure, sizeof, typeof, unsized, virtual, yield}  %
, morekeywords=[2]{Add, AddAssign, Any, AsciiExt, AsInner, AsInnerMut, AsMut, AsRawFd, AsRawHandle, AsRawSocket, AsRef, Binary, BitAnd, BitAndAssign, Bitor, BitOr, BitOrAssign, BitXor, BitXorAssign, Borrow, BorrowMut, Boxed, BoxPlace, BufRead, BuildHasher, CastInto, CharExt, Clone, CoerceUnsized, CommandExt, Copy, Debug, DecodableFloat, Default, Deref, DerefMut, DirBuilderExt, DirEntryExt, Display, Div, DivAssign, DoubleEndedIterator, DoubleEndedSearcher, Drop, EnvKey, Eq, Error, ExactSizeIterator, ExitStatusExt, Extend, FileExt, FileTypeExt, Float, Fn, FnBox, FnMut, FnOnce, Freeze, From, FromInner, FromIterator, FromRawFd, FromRawHandle, FromRawSocket, FromStr, FullOps, FusedIterator, Generator, Hash, Hasher, Index, IndexMut, InPlace, Int, Into, IntoCow, IntoInner, IntoIterator, IntoRawFd, IntoRawHandle, IntoRawSocket, IsMinusOne, IsZero, Iterator, JoinHandleExt, LargeInt, LowerExp, LowerHex, MetadataExt, Mul, MulAssign, Neg, Not, Octal, OpenOptionsExt, Ord, OsStrExt, OsStringExt, Packet, PartialEq, PartialOrd, Pattern, PermissionsExt, Place, Placer, Pointer, Product, Put, RangeArgument, RawFloat, Read, Rem, RemAssign, Seek, Shl, ShlAssign, Shr, ShrAssign, Sized, SliceConcatExt, SliceExt, SliceIndex, Stats, Step, StrExt, Sub, SubAssign, Sum, Sync, TDynBenchFn, Terminal, Termination, ToOwned, ToSocketAddrs, ToString, Try, TryFrom, TryInto, UnicodeStr, Unsize, UpperExp, UpperHex, WideInt, Write}
, morekeywords=[2]{Send}  %
, morekeywords=[3]{bool, char, f32, f64, i8, i16, i32, i64, isize, str, u8, u16, u32, u64, unit, usize, i128, u128}  %
, morekeywords=[4]{Err, false, None, Ok, Some, true}  %
, morekeywords=[3]{AccessError, Adddf3, AddI128, AddoI128, AddoU128, ADDRESS, ADDRESS64, addrinfo, ADDRINFOA, AddrParseError, Addsf3, AddU128, advice, aiocb, Alignment, AllocErr, AnonPipe, Answer, Arc, Args, ArgsInnerDebug, ArgsOs, Argument, Arguments, ArgumentV1, Ashldi3, Ashlti3, Ashrdi3, Ashrti3, AssertParamIsClone, AssertParamIsCopy, AssertParamIsEq, AssertUnwindSafe, AtomicBool, AtomicPtr, Attr, auxtype, auxv, BackPlace, BacktraceContext, Barrier, BarrierWaitResult, Bencher, BenchMode, BenchSamples, BinaryHeap, BinaryHeapPlace, blkcnt, blkcnt64, blksize, BOOL, boolean, BOOLEAN, BoolTrie, BorrowError, BorrowMutError, Bound, Box, bpf, BTreeMap, BTreeSet, Bucket, BucketState, Buf, BufReader, BufWriter, Builder, BuildHasherDefault, BY, BYTE, Bytes, CannotReallocInPlace, cc, Cell, Chain, CHAR, CharIndices, CharPredicateSearcher, Chars, CharSearcher, CharsError, CharSliceSearcher, CharTryFromError, Child, ChildPipes, ChildStderr, ChildStdin, ChildStdio, ChildStdout, Chunks, ChunksMut, ciovec, clock, clockid, Cloned, cmsgcred, cmsghdr, CodePoint, Color, ColorConfig, Command, CommandEnv, Component, Components, CONDITION, condvar, Condvar, CONSOLE, CONTEXT, Count, Cow, cpu, CRITICAL, CStr, CString, CStringArray, Cursor, Cycle, CycleIter, daddr, DebugList, DebugMap, DebugSet, DebugStruct, DebugTuple, Decimal, Decoded, DecodeUtf16, DecodeUtf16Error, DecodeUtf8, DefaultEnvKey, DefaultHasher, dev, device, Difference, Digit32, DIR, DirBuilder, dircookie, dirent, dirent64, DirEntry, Discriminant, DISPATCHER, Display, Divdf3, Divdi3, Divmoddi4, Divmodsi4, Divsf3, Divsi3, Divti3, dl, Dl, Dlmalloc, Dns, DnsAnswer, DnsQuery, dqblk, Drain, DrainFilter, Dtor, Duration, DwarfReader, DWORD, DWORDLONG, DynamicLibrary, Edge, EHAction, EHContext, Elf32, Elf64, Empty, EmptyBucket, EncodeUtf16, EncodeWide, Entry, EntryPlace, Enumerate, Env, epoll, errno, Error, ErrorKind, EscapeDebug, EscapeDefault, EscapeUnicode, event, Event, eventrwflags, eventtype, ExactChunks, ExactChunksMut, EXCEPTION, Excess, ExchangeHeapSingleton, exit, exitcode, ExitStatus, Failure, fd, fdflags, fdsflags, fdstat, ff, fflags, File, FILE, FileAttr, filedelta, FileDesc, FilePermissions, filesize, filestat, FILETIME, filetype, FileType, Filter, FilterMap, Fixdfdi, Fixdfsi, Fixdfti, Fixsfdi, Fixsfsi, Fixsfti, Fixunsdfdi, Fixunsdfsi, Fixunsdfti, Fixunssfdi, Fixunssfsi, Fixunssfti, Flag, FlatMap, Floatdidf, FLOATING, Floatsidf, Floatsisf, Floattidf, Floattisf, Floatundidf, Floatunsidf, Floatunsisf, Floatuntidf, Floatuntisf, flock, ForceResult, FormatSpec, Formatted, Formatter, Fp, FpCategory, fpos, fpos64, fpreg, fpregset, FPUControlWord, Frame, FromBytesWithNulError, FromUtf16Error, FromUtf8Error, FrontPlace, fsblkcnt, fsfilcnt, fsflags, fsid, fstore, fsword, FullBucket, FullBucketMut, FullDecoded, Fuse, GapThenFull, GeneratorState, gid, glob, glob64, GlobalDlmalloc, greg, group, GROUP, Guard, GUID, Handle, HANDLE, Handler, HashMap, HashSet, Heap, HINSTANCE, HMODULE, hostent, HRESULT, id, idtype, if, ifaddrs, IMAGEHLP, Immut, in, in6, Incoming, Infallible, Initializer, ino, ino64, inode, input, InsertResult, Inspect, Instant, int16, int32, int64, int8, integer, IntermediateBox, Internal, Intersection, intmax, IntoInnerError, IntoIter, IntoStringError, intptr, InvalidSequence, iovec, ip, IpAddr, ipc, Ipv4Addr, ipv6, Ipv6Addr, Ipv6MulticastScope, Iter, IterMut, itimerspec, itimerval, jail, JoinHandle, JoinPathsError, KDHELP64, kevent, kevent64, key, Key, Keys, KV, l4, LARGE, lastlog, launchpad, Layout, Lazy, lconv, Leaf, LeafOrInternal, Lines, LinesAny, LineWriter, linger, linkcount, LinkedList, load, locale, LocalKey, LocalKeyState, Location, lock, LockResult, loff, LONG, lookup, lookupflags, LookupHost, LPBOOL, LPBY, LPBYTE, LPCSTR, LPCVOID, LPCWSTR, LPDWORD, LPFILETIME, LPHANDLE, LPOVERLAPPED, LPPROCESS, LPPROGRESS, LPSECURITY, LPSTARTUPINFO, LPSTR, LPVOID, LPWCH, LPWIN32, LPWSADATA, LPWSAPROTOCOL, LPWSTR, Lshrdi3, Lshrti3, lwpid, M128A, mach, major, Map, mcontext, Metadata, Metric, MetricMap, mflags, minor, mmsghdr, Moddi3, mode, Modsi3, Modti3, MonitorMsg, MOUNT, mprot, mq, mqd, msflags, msghdr, msginfo, msglen, msgqnum, msqid, Muldf3, Mulodi4, Mulosi4, Muloti4, Mulsf3, Multi3, Mut, Mutex, MutexGuard, MyCollection, n16, NamePadding, NativeLibBoilerplate, nfds, nl, nlink, NodeRef, NoneError, NonNull, NonZero, nthreads, NulError, OccupiedEntry, off, off64, oflags, Once, OnceState, OpenOptions, Option, Options, OptRes, Ordering, OsStr, OsString, Output, OVERLAPPED, Owned, Packet, PanicInfo, Param, ParseBoolError, ParseCharError, ParseError, ParseFloatError, ParseIntError, ParseResult, Part, passwd, Path, PathBuf, PCONDITION, PCONSOLE, Peekable, PeekMut, Permissions, PhantomData, pid, Pipes, PlaceBack, PlaceFront, PLARGE, PoisonError, pollfd, PopResult, port, Position, Powidf2, Powisf2, Prefix, PrefixComponent, PrintFormat, proc, Process, PROCESS, processentry, protoent, PSRWLOCK, pthread, ptr, ptrdiff, PVECTORED, Queue, radvisory, RandomState, Range, RangeFrom, RangeFull, RangeInclusive, RangeMut, RangeTo, RangeToInclusive, RawBucket, RawFd, RawHandle, RawPthread, RawSocket, RawTable, RawVec, Rc, ReadDir, Receiver, recv, RecvError, RecvTimeoutError, ReentrantMutex, ReentrantMutexGuard, Ref, RefCell, RefMut, REPARSE, Repeat, Result, Rev, Reverse, riflags, rights, rlim, rlim64, rlimit, rlimit64, roflags, Root, RSplit, RSplitMut, RSplitN, RSplitNMut, RUNTIME, rusage, RwLock, RWLock, RwLockReadGuard, RwLockWriteGuard, sa, SafeHash, Scan, sched, scope, sdflags, SearchResult, SearchStep, SECURITY, SeekFrom, segment, Select, SelectionResult, sem, sembuf, send, Sender, SendError, servent, sf, Shared, shmatt, shmid, ShortReader, ShouldPanic, Shutdown, siflags, sigaction, SigAction, sigevent, sighandler, siginfo, Sign, signal, signalfd, SignalToken, sigset, sigval, Sink, SipHasher, SipHasher13, SipHasher24, size, SIZE, Skip, SkipWhile, Slice, SmallBoolTrie, sockaddr, SOCKADDR, sockcred, Socket, SOCKET, SocketAddr, SocketAddrV4, SocketAddrV6, socklen, speed, Splice, Split, SplitMut, SplitN, SplitNMut, SplitPaths, SplitWhitespace, spwd, SRWLOCK, ssize, stack, STACKFRAME64, StartResult, STARTUPINFO, stat, Stat, stat64, statfs, statfs64, StaticKey, statvfs, StatVfs, statvfs64, Stderr, StderrLock, StderrTerminal, Stdin, StdinLock, Stdio, StdioPipes, Stdout, StdoutLock, StdoutTerminal, StepBy, String, StripPrefixError, StrSearcher, subclockflags, Subdf3, SubI128, SuboI128, SuboU128, subrwflags, subscription, Subsf3, SubU128, Summary, suseconds, SYMBOL, SYMBOLIC, SymmetricDifference, SyncSender, sysinfo, System, SystemTime, SystemTimeError, Take, TakeWhile, tcb, tcflag, TcpListener, TcpStream, TempDir, TermInfo, TerminfoTerminal, termios, termios2, TestDesc, TestDescAndFn, TestEvent, TestFn, TestName, TestOpts, TestResult, Thread, threadattr, threadentry, ThreadId, tid, time, time64, timespec, TimeSpec, timestamp, timeval, timeval32, timezone, tm, tms, ToLowercase, ToUppercase, TraitObject, TryFromIntError, TryFromSliceError, TryIter, TryLockError, TryLockResult, TryRecvError, TrySendError, TypeId, U64x2, ucontext, ucred, Udivdi3, Udivmoddi4, Udivmodsi4, Udivmodti4, Udivsi3, Udivti3, UdpSocket, uid, UINT, uint16, uint32, uint64, uint8, uintmax, uintptr, ulflags, ULONG, ULONGLONG, Umoddi3, Umodsi3, Umodti3, UnicodeVersion, Union, Unique, UnixDatagram, UnixListener, UnixStream, Unpacked, UnsafeCell, UNWIND, UpgradeResult, useconds, user, userdata, USHORT, Utf16Encoder, Utf8Error, Utf8Lossy, Utf8LossyChunk, Utf8LossyChunksIter, utimbuf, utmp, utmpx, utsname, uuid, VacantEntry, Values, ValuesMut, VarError, Variables, Vars, VarsOs, Vec, VecDeque, vm, Void, WaitTimeoutResult, WaitToken, wchar, WCHAR, Weak, whence, WIN32, WinConsole, Windows, WindowsEnvKey, winsize, WORD, Wrapping, wrlen, WSADATA, WSAPROTOCOL, WSAPROTOCOLCHAIN, Wtf8, Wtf8Buf, Wtf8CodePoints, xsw, xucred, Zip, zx}
, morekeywords=[5]{assert!, assert_eq!, assert_ne!, cfg!, column!, compile_error!, concat!, concat_idents!, debug_assert!, debug_assert_eq!, debug_assert_ne!, env!, eprint!, eprintln!, file!, format!, format_args!, include!, include_bytes!, include_str!, line!, module_path!, option_env!, panic!, print!, println!, select!, stringify!, thread_local!, try!, unimplemented!, unreachable!, vec!, write!, writeln!}  %
}%

\usetikzlibrary{calc,tikzmark}

\newcommand{\codebox}[2]{%
  \begin{tikzpicture}[remember picture,overlay]
    \draw[rounded corners=2pt, line width=0.6pt, dashed]
      ([xshift=-0.3em,yshift=1.5ex]pic cs:#1) rectangle
      ([xshift= 0.6em,yshift=-0.6ex]pic cs:#2);
  \end{tikzpicture}%
}

\makeatletter
\renewcommand{\nllabel}[1]
 {{\let\@currentlabel\algocf@currentlabel
  \let\@currentcounter\algocf@currentcounter
  \label{#1}}}%

\renewcommand{\algocf@nl@sethref}[1]{%
  \renewcommand{\theHAlgoLine}{\thealgocfproc.#1}%
  \hyper@refstepcounter{AlgoLine}%
  \gdef\algocf@currentlabel{#1}%
  \gdef\algocf@currentcounter{AlgoLine}%
 }%
\makeatother

\usepackage{setspace}

\SetKwFunction{AlgMain}{main}
\SetKwArray{AlgCorpus}{corpus}
\SetKwArray{AlgInstSeqs}{instSeqs}

\SetKwFunction{AlgExtract}{Extract}
\SetKwFunction{AlgExtractFromBB}{ExtractSeqsFromBB}
\SetKwFunction{AlgEncode}{Hash}
\SetKwFunction{AlgConvertToFunction}{WrapAsFunc}
\SetKwFunction{AlgInvokeLLM}{InvokeLLM}
\SetKwFunction{AlgRunOPT}{RunOPT}
\SetKwFunction{AlgIsInteresting}{IsInteresting}
\SetKwFunction{AlgRunAlive}{RunAlive2}

\SetKwData{AlgModule}{module}
\SetKwData{AlgFunction}{function}
\SetKwData{AlgBasicBlock}{bb}
\SetKwData{AlgInstruction}{inst}
\SetKwData{AlgUsed}{added}
\SetKwData{AlgDedupSet}{dedup\_set}
\SetKwData{AlgEncoding}{digest}
\SetKwData{AlgConvertedFunction}{wrapped\_func}
\SetKwData{AlgCandidate}{candidate}
\SetKwData{AlgOPTResult}{result\_OPT}
\SetKwData{AlgAliveResult}{result\_Alive2}
\SetKwData{AlgIsCorrect}{is\_correct}
\SetKwData{AlgCounterExample}{counter\_example}
\SetKwData{AlgNewCandidate}{new\_candidate}
\SetKwData{AlgErrorMessage}{error\_message}
\SetKwData{AlgAttemptCounter}{counter}
\SetKwData{AlgAttemptLimit}{ATTEMPT\_LIMIT}
\SetKwData{AlgFailed}{is\_failed}
\SetKwData{AlgFeedback}{feedback}

\SetKwArray{AlgUniqueInstSeqSet}{result}

\SetKwArray{AlgInstSeqList}{seq\_set}
\SetKwArray{AlgNewSeqList}{new\_set}

\SetKwData{AlgInstSeq}{instSeq}
\SetKwData{AlgNewInstSeq}{newInstSeq}

\usepackage{cleveref} %

\Crefname{algocf}{Algorithm}{Algorithms}
\crefname{algocf}{Algorithm}{Algorithms}

\Crefname{algorithm}{Algorithm}{Algorithms}
\crefname{algorithm}{Algorithm}{Algorithms}

\crefname{appendix}{Appendix}{Appendices}
\Crefname{appendix}{Appendix}{Appendices}

\Crefname{figure}{Figure}{Figures}
\crefname{figure}{Figure}{Figures}

\crefname{listing}{Listing}{Listings}
\Crefname{listing}{Listing}{Listings}

\Crefname{table}{Table}{Tables}
\crefname{table}{Table}{Tables}

\crefname{thm}{Theorem}{Theorems}
\Crefname{thm}{Theorem}{Theorems}

\crefname{equation}{Equation}{Equations}
\Crefname{equation}{Equation}{Equations}

\crefformat{chapter}{\S~#2#1#3}
\crefmultiformat{chapter}{\S\S~#2#1#3}{ and~#2#1#3}{, #2#1#3}{, and~#2#1#3}

\crefformat{section}{\S~#2#1#3}
\crefmultiformat{section}{\S\S~#2#1#3}{ and~#2#1#3}{, #2#1#3}{, and~#2#1#3}

\newif\ifInCopypaper
\InCopypaperfalse

\AtBeginDocument{%
  \providecommand\BibTeX{{%
    \normalfont B\kern-0.5em{\scshape i\kern-0.25em b}\kern-0.8em\TeX}}}

\copyrightyear{2026}
\acmYear{2026}
\setcopyright{cc}
\setcctype{by}
\acmConference[ASPLOS '26]{Proceedings of the 31st ACM International Conference on Architectural Support for Programming Languages and Operating Systems, Volume 2}{March 22--26, 2026}{Pittsburgh, PA, USA}
\acmBooktitle{Proceedings of the 31st ACM International Conference on Architectural Support for Programming Languages and Operating Systems, Volume 2 (ASPLOS '26), March 22--26, 2026, Pittsburgh, PA, USA}
\acmPrice{}
\acmDOI{10.1145/3779212.3790184}
\acmISBN{979-8-4007-2359-9/2026/03}

\begin{document}

\title{LPO: Discovering Missed Peephole Optimizations with Large Language Models}

\author{Zhenyang Xu}
\authornote{Zhenyang Xu and Hongxu Xu contributed equally to this work.}
\affiliation{%
	\institution{Cheriton School of Computer Science, University of Waterloo}
	\city{Waterloo}
	\country{Canada}}
\email{zhenyang.xu@uwaterloo.ca}

\author{Hongxu Xu}
\authornotemark[1]
\affiliation{%
	\institution{Cheriton School of Computer Science, University of Waterloo}
	\city{Waterloo}
	\country{Canada}}
\email{hongxu.xu@uwaterloo.ca}

\author{Yongqiang Tian}
\affiliation{%
	\institution{Department of Software Systems \& Cybersecurity, Monash University}
	\city{Melbourne}
	\country{Australia}}
\email{yongqiang.tian@monash.edu}

\author{Xintong Zhou}
\affiliation{%
	\institution{Cheriton School of Computer Science, University of Waterloo}
	\city{Waterloo}
	\country{Canada}}
\email{x27zhou@uwaterloo.ca}

\author{Chengnian Sun}
\affiliation{%
	\institution{Cheriton School of Computer Science, University of Waterloo}
	\city{Waterloo}
	\country{Canada}}
\email{cnsun@uwaterloo.ca}

\begin{abstract}

Peephole optimization is an essential class of compiler optimizations
that targets small, inefficient instruction sequences within programs.
By replacing such suboptimal instructions with refined and
more optimal sequences, these optimizations not only
directly optimize code size and  performance,
but also enable more
transformations in the subsequent optimization pipeline.
Despite their importance,
discovering new and effective peephole optimizations remains challenging
due to the complexity and breadth of instruction sets.
Prior approaches either lack scalability or have significant restrictions
on the peephole optimizations that they can find.

This paper  introduces \proj,
a novel automated framework
to discover missed peephole optimizations.
Our key insight is that,
Large Language Models (\llms)
are effective at creative exploration but susceptible to hallucinations;
conversely, formal verification techniques provide rigorous guarantees
but struggle with creative discovery.
By synergistically combining the strengths of \llms and formal verifiers in a closed-loop feedback mechanism,
\proj can effectively discover verified peephole
optimizations that were previously missed.

We comprehensively evaluated \proj within
\llvm ecosystems.
Our evaluation shows that \proj can successfully identify up to
\numberOfBenchmarksLLMsCanDetect out of \numberOfBenchmarks
previously reported missed optimizations in \llvm.
In contrast, the recently proposed superoptimizers for \llvm, \souper and \minotaur
detected \numberOfBenchmarksFormalCanDetect and \numberOfBenchmarksMinotaurCanDetect of them, respectively.
More importantly, within \monthsOfExperiment months of development and intermittent testing, \proj
found \reportedInTotal missed peephole optimizations,
of which \confirmedBugs were confirmed and an additional \fixedBugs had already been fixed in \llvm.
These results demonstrate \proj's strong potential to continuously
uncover new optimizations as \llms' reasoning improves.

\end{abstract}



\maketitle

\section{Introduction}
\label{sec:intro}
Compiler optimizations are a cornerstone of modern software development,
designed to enhance a program's performance without altering its behavior~\cite{dragonBook}.
A large portion of optimizations are applied in a series of modular stages called
compiler optimization passes, which analyze and transform a program's
intermediate representation (\ir). Among these, peephole optimization is a
critical technique that performs local transformations on small, suboptimal
instruction sequences. It operates by identifying a ``window'' (or ``peephole'') of dependent
instructions and replacing them with a potentially more efficient and refined
(\ie, equivalent or with less non-determinism when the original sequence has undefined behavior)
sequence.
This process is essential for tasks like eliminating
redundant code, performing constant folding, and simplifying complex
operations. Importantly, these local improvements can also enable more
significant optimizations by subsequent passes, making peephole
optimization a foundational and indispensable technique in compiler design.

Despite its importance, the development of a comprehensive, effective peephole
optimizer remains a significant challenge.
Unlike other optimizations, such
as dead code elimination or loop-invariant code motion,
which may be
implemented with a single, unified algorithm, peephole optimizations are a
collection of pattern-matching and rewriting rules. The complexity and
vastness of modern instruction sets cause the number of potential
optimization patterns to multiply rapidly,
making it exceedingly difficult to
develop a complete optimizer. This is why optimizers in widely used
compilers like \llvm are continuously evolving, with new patterns being
generalized or added~\cite{menendez2016termination,instcombinepr}.

Previous approaches for discovering new peephole optimizations can be classified into three categories:

\noindent\underline{Manual Inspection:} A straightforward way to discover new peephole optimization
is manually inspecting the generated \ir or assembly code to identify suboptimal
instruction patterns.
This approach has led to the discovery of a considerable number of
optimizations; however, it is labor-intensive, requires deep domain
expertise, and has limited scalability.

\noindent\underline{Differential Testing:} Prior studies have proposed using
differential testing to automatically
detect missed optimization opportunities~\cite{theodoridis2022finding,liu2023exploring,barany2018finding,italiano2024finding}.
This is typically performed in one of the two ways:
by identifying differences among the results of
different optimizing compilers,
or by comparing the compilation results of a pair of semantically equivalent programs.
While these approaches have proven effective,
they are limited to discovering optimizations
that have already been implemented in other compilers
or that are not performing as expected in certain cases,
rather than discovering entirely new classes of missed optimizations.

\noindent\underline{Superoptimization:} Given a sequence of instructions,
a superoptimizer aims to find the optimal instruction sequence through
stochastic search or program synthesis~\cite{sasnauskas2017souper,schkufza2013stochastic,
joshi2002denali,bansal2006automatic,buchwald2015optgen,DBLP:conf/osdi/BansalA08,liu2024minotaur}.
The resulting transformation can then be generalized
and implemented as a peephole optimization.
However, these approaches are typically computationally expensive,
which constrains the code size
they can process
and limits them to
supporting only a subset of the instruction set.
For example, \souper, a synthesized superoptimizer for \llvm \ir,
does not support memory, floating-point, or vector instructions.

\myparagraph{\proj}
This paper explores the feasibility of using Large Language Models (\llms)
to discover new peephole optimizations. While \llms are highly effective
at creative exploration and reasoning,
they are prone to generating incorrect or ``hallucinated'' answers.
Conversely, formal verification can guarantee the correctness of an optimization
but is often unsuited for the creative task of searching for new ones.
We propose \proj, a novel technique that resolves
this tension by combining the creative search capabilities
of \llms with the rigorous guarantees of formal verification.
The \proj framework operates as an automated discovery engine
for missed peephole optimizations.
Its workflow is built around three key components:
an extractor, an \llm-based optimizer, and a verifier.
First, the extractor analyzes programs
to identify candidate instruction sequences for optimization.
These candidates are then passed to the \llm-based optimizer,
which suggests a potentially more efficient equivalent.
The verifier then rigorously checks the proposed optimization
for refinement and performance improvement.
If the optimization is verified, it is saved for further investigation.
If the verification fails,
the verifier's output is used as targeted feedback to the \llm, guiding the \llm
to correct its proposal in a new attempt.
This closed-loop process allows us to
systematically and automatically find new, verified peephole optimizations.

We thoroughly evaluated \proj's ability to find new peephole optimizations.
We first curated a benchmark suite of \numberOfBenchmarks
missed peephole optimizations
that were recently reported in the \llvm repository.
Our results show that \proj can identify
up to \numberOfBenchmarksLLMsCanDetect of them,
while \souper and \minotaur, two recently proposed synthesizing superoptimizers,
detected \numberOfBenchmarksFormalCanDetect and \numberOfBenchmarksMinotaurCanDetect.
Furthermore, we tested \proj's capability to discover
unreported, missed optimizations.
Within \monthsOfExperiment months of development and intermittent testing,
\proj
found \reportedInTotal missed peephole optimizations,
with \confirmedBugs confirmed and an additional \fixedBugs already fixed in \llvm.
Notably, \souper and \minotaur fail to identify \confirmedSouperCannot and \confirmedMinotaurCannot
of these confirmed or fixed cases, respectively.
These results clearly demonstrate the strong potential
of our approach,
which combines the creative search of \llms
with the rigorous correctness of formal verification,
to continuously find new optimizations as the reasoning abilities of \llms advance.

\myparagraph{Contributions} We make the following contributions.
\begin{itemize}[leftmargin=1em, topsep=0pt, parsep=0pt]

\item We present \proj,
a novel framework that synergistically combines the creative search capabilities of \llms with the rigorous guarantees of formal verification,
providing an automated solution
to the long-standing challenge of discovering new peephole optimizations.

\item We introduce a new methodology for compiler optimization discovery
that leverages \llms to overcome the limitations of prior approaches
like manual inspection, differential testing, and traditional
superoptimization.
A key innovation of our method is a closed-loop feedback mechanism, where formal verification results are used to iteratively guide the \llm's
search for correct optimizations.

\item
We comprehensively evaluated \proj in the \llvm ecosystem
and showed that it can uncover previously unreported missed optimizations.
\proj identified \reportedInTotal missed peephole
optimizations, with \confirmedBugs confirmed and \fixedBugs already
fixed in \llvm.
Notably, \confirmedSouperCannot and \confirmedMinotaurCannot of these optimizations
cannot be detected by \souper and \minotaur,
two synthesizing superoptimizers for \llvm, respectively.
These results highlight the advantages of \proj over prior approaches and
establish it as a validated, practical tool for compiler development.

\item
To ensure reproducibility and facilitate future research, we have
released the replication package and experimental data at \toolurl.

\end{itemize}

\section{Background}
\label{sec:background}
We built \proj on top of \llvm
and used \proj to find missed peephole optimizations within \llvm.
We chose \llvm because of its flexible and user-friendly tool-chain.
While our implementation targets \llvm, the core concept of \proj is general and
can be applied to other compilers.
This section introduces necessary background knowledge on
\llvm \ir,
peephole optimizations,
and translation validation.

\begin{figure*}[t]
  \begin{minipage}[c]{0.49\linewidth}
    \begin{subfigure}[c]{\linewidth}
            \begin{lstlisting}[language=rust,
                numbers=none, xleftmargin=0em,numberblanklines,
                showstringspaces=false, stringstyle=\small\ttfamily, escapechar=|]
pub fn clamp(inp: &[i32], out: &mut [u8]) {
    for (&i, o) in inp.iter().zip(out.iter_mut()) {
        *o = i.clamp(0, 255) as u8;
    }
}
            \end{lstlisting}
            \caption{A \texttt{clamp} function that clamps
                a sequence of signed 32-bit integers to unsigned 8-bit integers.
                }
            \label{fig:rust-clamp}
    \end{subfigure}

\vspace*{9px}

\begin{subfigure}[c]{\linewidth}
\begin{lstlisting}[language=llvm,numbers=none, xleftmargin=0em,numberblanklines,
    showstringspaces=false, stringstyle=\small\ttfamily]
define i8 @src(i32 %0) {
  %2 = icmp slt i32 %0, 0
  %3 = tail call i32 @llvm.umin.i32(i32 %0, i32 255)
  %4 = trunc nuw i32 %3 to i8
  %5 = select i1 %2, i8 0, i8 %4
  ret i8 %5
}
\end{lstlisting}
\caption{The suboptimal instruction sequence.}
\label{subfig:alive2-input-src}
\end{subfigure}

\vspace*{9px}

\begin{subfigure}[c]{\linewidth}
\begin{lstlisting}[language=llvm,numbers=none, xleftmargin=0em,numberblanklines,
    showstringspaces=false, stringstyle=\small\ttfamily]
define i8 @tgt(i32 %0) {
  %2 = tail call i32 @llvm.smax.i32(i32 %0, i32 0)
  %3 = tail call i32 @llvm.umin.i32(i32 %2, i32 255)
  %4 = trunc nuw i32 %3 to i8
  ret i8 %4
}
\end{lstlisting}
\caption{The expected optimal instruction sequence.}
\label{subfig:alive2-input-tgt}
\end{subfigure}
\end{minipage}
\hfill
\begin{minipage}[c]{0.49\linewidth}
\begin{subfigure}[c]{\linewidth}
\begin{lstlisting}[language=llvm,
    numbers=left,xleftmargin=3em,numberblanklines,
    basicstyle=\scriptsize\ttfamily,
    showstringspaces=false, stringstyle=\small\ttfamily]
(*@\dots@*)
define void @clamp((*@\dots@*)){
  (*@\dots@*)
vector.body:(*@\label{line:clamp-llvm-ir:vector.body}@*)
  %i = phi i64 [0,%vector.ph],[%i.next,%vector.body]
  %0 = getelementptr inbounds nuw i32,ptr %inp,i64 %i
  %1 = getelementptr inbounds nuw i8,ptr %out,i64 %i
  %2 = getelementptr inbounds nuw i8,ptr %0,i64 16
  %wide.load = load <4 x i32>,ptr %0,align 4
  %wide.load7 = load <4 x i32>,ptr %2,align 4
  (*@\tikzmark{S}@*)%3 = icmp slt <4 x i32> %wide.load,zeroinitializer
  %4 = icmp slt <4 x i32> %wide.load7,zeroinitializer
  %5 = tail call <4 x i32> @llvm.umin.v4i32(
    <4 x i32> %wide.load, <4 x i32> splat (i32 255) )
  %6 = tail call <4 x i32> @llvm.umin.v4i32(
    <4 x i32> %wide.load7, <4 x i32> splat (i32 255) )
  %7 = trunc nuw <4 x i32> %5 to <4 x i8>
  %8 = trunc nuw <4 x i32> %6 to <4 x i8>
  %9 = select <4 x i1> %3,
    <4 x i8> zeroinitializer, <4 x i8> %7
  %10 = select <4 x i1> %4,
    <4 x i8> zeroinitializer, <4 x i8> %8            (*@\tikzmark{E}@*)
  %11 = getelementptr inbounds nuw i8, ptr %1, i64 4
  store <4 x i8> %9, ptr %1, align 1
  store <4 x i8> %10, ptr %11, align 1
  %i.next = add nuw i64 %i, 8
  %12 = icmp eq i64 %i.next, %n.vec
  br i1 %12, label %middle.block, label %vector.body
middle.block:
  (*@\dots@*)
} (*@\dots@*)
\end{lstlisting}
\codebox{S}{E}
\caption{The \llvm IR module of the \texttt{clamp} function in \cref{fig:rust-clamp},
    with the suboptimal instruction sequence highlighted by the dashed box.
}
\label{fig:llvm-ir}
\end{subfigure}
\end{minipage}
\caption{
    (\subref{fig:rust-clamp}) and (\subref{fig:llvm-ir})
    show a Rust function \texttt{clamp} revealing a missed optimization in \llvm, and
    its corresponding \llvm \ir module.
    (\subref{subfig:alive2-input-src}) and (\subref{subfig:alive2-input-tgt})
    are a pair of simplified \llvm \ir functions
    illustrate the essence of the optimization.
}
\label{fig:rust-clamp-llvm-ir}

\end{figure*}

\subsection{\llvm Intermediate Representation}

\llvm
is an open-source compiler framework
for various programming
languages and architectures~\cite{lattner2002llvm,lattner2004llvm,llvm2025llvm}.
It has a low-level, typed \ir in
static single-assignment form, designed for expressiveness and
extensibility.

\llvm \ir programs consist of \textit{modules}.
Each module is a translation unit of the input program,
and also is the top level container of all other IR objects,
such as functions and global variables~\cite{llvm2025languageref}.
\Cref{fig:llvm-ir} illustrates an \llvm \ir
module
(irrelevant instructions, attributes and declarations are hidden for clarity)
that contains a function
generated from the original \rust function \clamp
in \cref{fig:rust-clamp}.
The \ir function is organized into basic blocks, such as
\texttt{vector.body} (line~\ref{line:clamp-llvm-ir:vector.body})
shown in the figure.
Each basic block comprises a sequence of instructions,
has a single entry point (optionally denoted by a label),
and ends with a terminator instruction that determines
which basic block will be executed next, or whether the function will return.
(\eg, \mycode{br} and \mycode{ret} instructions)~\cite{llvm2025languageref}.

\subsection{Peephole Optimization in \llvm}

\emph{Peephole optimizations} identify and replace inefficient patterns with
more efficient ones, by examining small instruction windows~\cite{mckeeman1965peephole}.
For instance, the condition $a - b > a + b$ can be optimized to $b < 0$.
Modern peephole optimizations are applied not only to assembly code,
but also at
higher compilation levels, including abstract syntax tree
and \ir~\cite{fischer1991crafting}.

The \instcombine pass in \llvm is the primary peephole
optimization pass~\cite{mukherjee2024hydra,menendez2016termination},
which applies algebraic simplifications by
pattern matching and replacing instruction sequences with more efficient
alternatives~\cite{lopes2015provably}.
\instcombine also performs \emph{canonicalization},
transforming instructions into a consistent form.
For example, binary operators
with constant operands are rewritten to place the constant on the right-hand side.
Such canonicalization reduces the pattern search space and facilitates further
optimizations.

However, the development of a comprehensive and effective peephole optimizer is
extremely challenging~\cite{llvm2025instcombine,davidson1984automatic}.
Currently, the \llvm compiler contains over 25,000 lines of C++ code dedicated to peephole optimizations,
and this number is still growing with new patterns added or existing patterns being
generalized~\cite{menendez2016termination,instcombinepr}.

\myparagraph{Missed Optimization} The Rust program in \cref{fig:rust-clamp} and
its corresponding \llvm \ir in \cref{fig:llvm-ir} demonstrate a
reported missed peephole optimization.
The instruction sequence highlighted in the rectangle is suboptimal and can be further optimized.
After analysis and simplification, the missed optimization can be illustrated with a pair of minimized \llvm functions
as shown in \cref{subfig:alive2-input-src,subfig:alive2-input-tgt}.
In this example, the \texttt{src} function clamps the input using
\texttt{x < 0 ? 0 : umin(x, 255)}, whereas the \texttt{tgt} function achieves
the same effect with \texttt{umin(smax(x, 0), 255)}. This transformation
reduces the instruction count,
resulting in a more efficient implementation.

\subsection{Superoptimization}
\emph{Superoptimization} aims to transform instruction sequences to their optimal implementation
by searching the space of possible instructions~\cite{massalin1987superoptimizer}.
A superoptimizer can work either as an online optimizer that performs optimization during compilation,
or as an offline optimization generator that
helps identifying missed optimization.
As the computational cost of superoptimization grows exponentially
with the length of the instruction sequence
(\ie, the search space),
superoptimization is usually restricted within small code regions
and is often used for finding missed peephole optimizations~\cite{bansal2006automatic,buchwald2015optgen,sasnauskas2017souper}.

\myparagraph{\souper}
\souper~\cite{sasnauskas2017souper} is a superoptimizer for \llvm \ir
to identify more efficient integer-typed instruction sequences by extracting
functions from a module and traversing dataflow edges backward from the return
instruction. Unlike exhaustive search, \souper employs counterexample-guided
synthesis to accelerate the discovery of optimal replacements. However, its
applicability is limited, as it does not support memory accesses,
floating-point, or vector instructions.

\subsection{Translation Validation}

\emph{Translation validation} is a formal technique for verifying that
the target code produced by compilers or optimizers
correctly implements the source code~\cite{necula2000translation,pnueli1998translation,lopes2021alive2}.
An automatic translation validation process requires
a formal specification of ``correct implementation'', which is a
refinement relation~\cite{pnueli1998translation}.
For instance, the \llvm \ir function produced by a transformation pass,
such as the \instcombine pass,
should display a subset of the behaviors of the original IR function.
The transformation can only eliminate non-determinism by refining
undefined behaviors, without introducing new ones.
When no undefined behavior exists, the refinement relation
becomes an equivalence~\cite{lopes2021alive2}.

\myparagraph{\alive}
\emph{\alive}~\cite{lopes2021alive2} is an automated translation validation tool for \llvm.
Given a pair of \llvm functions, \alive verifies whether the transformation from the source function
to the target function is correct (\ie, whether the source is refined by the target).
For example, given the \llvm function pair shown in \cref{subfig:alive2-input-src,subfig:alive2-input-tgt},
\alive can prove that the transformation from \texttt{src} to \texttt{tgt} is correct.
Conversely,
if a transformation is proven to be incorrect, \alive provides a counterexample
that refutes the refinement relation between the source and the target.
\alive is widely used in the \llvm community
and is integrated into the \llvm development workflow~\cite{llvm2025instcombine}.

\section{Methodology}
\label{sec:method}
\begin{figure*}[htbp]
\centering
\includegraphics[width=0.95\linewidth]{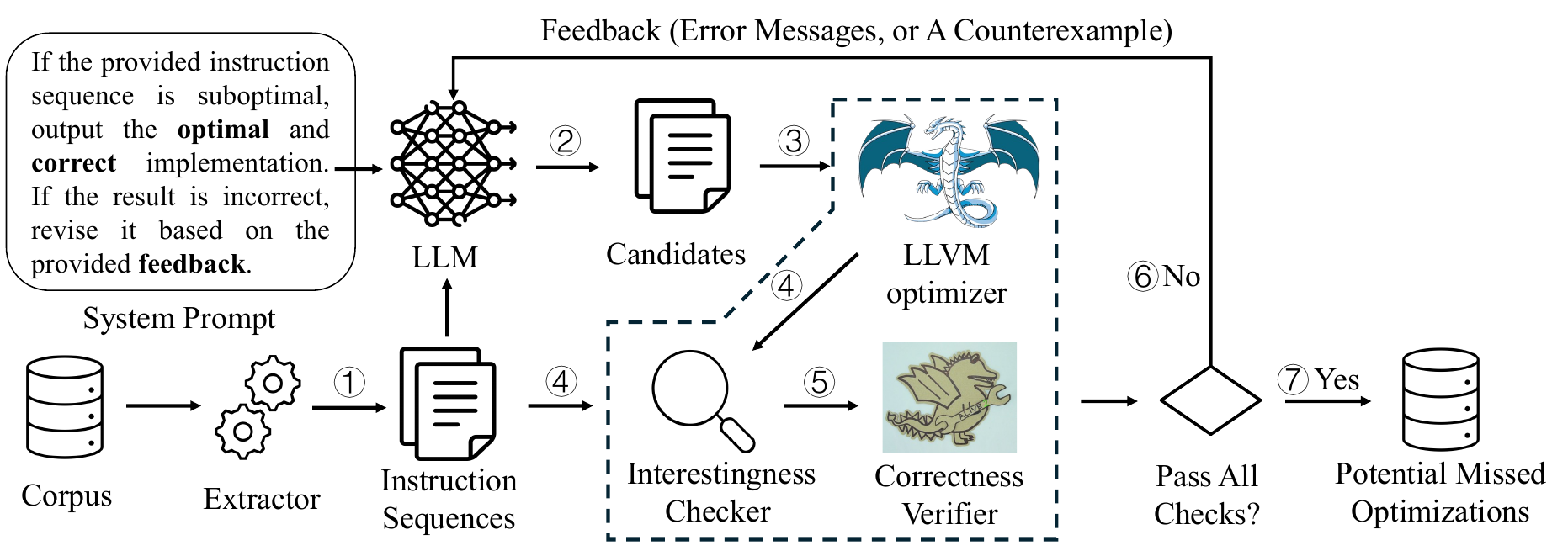}
\caption{The overall workflow of \proj. The verification phase is outlined by the dashed line.
}
\label{fig:overlall-workflow}
\end{figure*}

\cref{fig:overlall-workflow} and \cref{alg:main} show the overall workflow of \proj.

\myparagraph{Step \circled{1}: Extract instruction sequences}
Given a corpus of programs in \llvm IR that have been compiled and optimized by \llvm,
\proj first traverses the basic blocks in these IR programs
and extracts all the dependent instruction sequences within each basic block
(\crefrange{alg:main:extract:start}{alg:main:extract:end}).
The details about the extraction process is introduced in
\ifInCopypaper
\S 3.2.
\else
 \cref{subsec:extraction}.
\fi

\myparagraph{Step \circled{2}\circled{3}\circled{7}: Optimize the instruction sequence}
For each instruction sequence, \proj prompts an \llm to output the optimal instruction sequence
(\cref{alg:main:invoke:llm} and step \circled{2}).
For the candidate generate by the \llm,
\proj runs three checks to verify whether it potentially reveals a valid optimization
(\crefrange{alg:main:check:start}{alg:main:check:end}).
If all checks are passed, the instruction sequence and the corresponding candidate generated by the \llm
are saved for further analysis
(\cref{alg:main:save} and step \circled{7}).

\myparagraph{Step \circled{3}\circled{4}\circled{5}\circled{6}: Verify the candidate and provide feedback}
\llms are not always reliable.
The candidate produced by \llms can, for example,
(1) have syntax error,
(2) be sub-optimal,
(3) be not canonical,
(4) have no refinement relation to the original instruction sequence, or
(5) be not interesting as it is identical to the original sequence or is less efficient.
To tackle this limitation of \llms, \proj incorporates a verification phase
as outlined by the dashed line in \cref{fig:overlall-workflow}
to check all these mistakes that can be made by the \llm
(steps \circled{3}, \circled{4}, and \circled{5}).
First, the instruction sequence is sent to the \llvm optimizer, \opt.
If there is any syntax error, \proj uses the error message from \opt as
feedback and prompt the \llm to generate a new candidate
(\crefrange{alg:main:opt:fail:start}{alg:main:opt:fail:end}).
Otherwise, the optimized instruction sequence output by \opt becomes the
new candidate (\cref{alg:main:opt:update}).
Next, \proj checks whether the candidate is interesting, \ie,
whether the candidate potentially manifests a beneficial optimization
\ifInCopypaper
(more details in \S 3.3).
\else
(more details in \cref{subsec:verification}).
\fi
If the candidate is not interesting (\eg, is identical to the original instruction sequence),
\proj abandons this instruction sequence and moves to the next one
(\crefrange{alg:main:interesting:start}{alg:main:interesting:end}).
If the candidate is interesting, \proj further checks the correctness with \alive
(\cref{alg:main:alive}).
If the optimization is proven
to be incorrect, \proj uses
the counterexample output by \alive as the feedback
and prompts the \llm to output a correct version based on the feedback
(\crefrange{alg:main:alive:start}{alg:main:check:end}).
A parameter \AlgAttemptLimit can be configured to specify how many attempts
are allowed at most for optimizing each instruction sequence (\cref{alg:main:while:check}).
In our implementation, this parameter is set to 2.

The remainder of this section  details the workflow.
\ifInCopypaper
\S 3.1 llustrates the workflow with a concrete example.
\S 3.2 elaborates on how \proj extracts instruction sequences from the corpus.
\S 3.3 provides more details about the verification process.
\else
\cref{subsec:illustrative:example} illustrates the workflow with a concrete example.
\cref{subsec:extraction} elaborates on how \proj extracts instruction sequences from the corpus.
\cref{subsec:verification} details the verification process.
\fi

\begin{algorithm}[t]
\setstretch{0.85}
\small
\DontPrintSemicolon
\caption{%
    The overall workflow of \protect\proj
}
\label{alg:main}
\Input{\AlgCorpus, a list of \llvm \ir module}
\Output{A set of instruction sequence pairs that potentially reveal missed peephole optimizations}
\AlgInstSeqs $\gets \varnothing$ \; \nllabel{line:init:inst:seq}
\AlgDedupSet $\gets \varnothing$ \; \nllabel{line:init:dedup:set}
\ForEach{\AlgModule $\in$ \AlgCorpus} { \nllabel{alg:main:extract:start}
    \AlgInstSeqs $\gets$ \AlgInstSeqs $\cup$ \AlgExtract{\AlgModule, \AlgDedupSet} \;
} \nllabel{alg:main:extract:end}
\ForEach{\AlgInstSeq $\in$ \AlgInstSeqs} {
    \AlgAttemptCounter $\gets$ 0 \;
    \AlgFeedback $\gets$ empty string \;
    \While{\AlgAttemptCounter < \AlgAttemptLimit}{ \nllabel{alg:main:while:check}
        \AlgCandidate $\gets$ \AlgInvokeLLM{\AlgInstSeq, \AlgFeedback} \; \nllabel{alg:main:invoke:llm}
        \AlgOPTResult $\gets$ \AlgRunOPT{\AlgCandidate} \; \nllabel{alg:main:check:start}
        \If{\AlgOPTResult.\AlgFailed} { \nllabel{alg:main:opt:fail:start}
            \AlgAttemptCounter $\gets$ \AlgAttemptCounter + 1 \;
            \AlgFeedback $\gets$ \AlgOPTResult.\AlgErrorMessage \;
            \Continue \;
        } \nllabel{alg:main:opt:fail:end}
        \AlgCandidate $\gets$ \AlgOPTResult.\AlgNewCandidate \; \nllabel{alg:main:opt:update}
        \lIf{$\neg$\AlgIsInteresting{\AlgCandidate}} { \nllabel{alg:main:interesting:start}
            \Break \;
        } \nllabel{alg:main:interesting:end}
        \AlgAliveResult $\gets$ \AlgRunAlive{\AlgInstSeq, \AlgCandidate} \; \nllabel{alg:main:alive}
        \If{$\neg$\AlgAliveResult.\AlgIsCorrect} { \nllabel{alg:main:alive:start}
            \AlgAttemptCounter $\gets$ \AlgAttemptCounter + 1 \;
            \AlgFeedback $\gets$ \AlgAliveResult.\AlgCounterExample \;
            \Continue \;
        } \nllabel{alg:main:check:end}
        save \AlgInstSeq and \AlgCandidate for further analysis \; \nllabel{alg:main:save}
        \Break \;
    }
}
\end{algorithm}
\subsection{Illustrative Example}
\begin{figure*}[htbp]
\begin{minipage}[b]{0.45\linewidth}
\begin{subfigure}[b]{\linewidth}
\begin{lstlisting}[language=llvm,numbers=none, xleftmargin=0em,numberblanklines,
    showstringspaces=false, stringstyle=\small\ttfamily, basicstyle=\scriptsize\ttfamily]
define <4 x i8> @src(i64 %a0, ptr %a1) {
entry:
 %0 = getelementptr inbounds nuw i32,ptr %a1,i64 %a0
 %wide.load = load <4 x i32>, ptr %0, align 4
 (*@\tikzmark{S2}@*)%3 = icmp slt <4 x i32> %wide.load, zeroinitializer(*@\tikzmark{E2}@*)
 %5 = tail call <4 x i32> @llvm.umin.v4i32(
   <4 x i32> %wide.load, <4 x i32> splat (i32 255) )
 %7 = trunc nuw <4 x i32> %5 to <4 x i8>
 (*@\tikzmark{S3}@*)%9 = select <4 x i1> %3,
   <4 x i8> zeroinitializer, <4 x i8> %7(*@\tikzmark{E3}@*)
 ret <4 x i8> %9
}
\end{lstlisting}
\caption{An instruction sequence (wrapped as a function)
extracted by \proj from the \llvm \ir module shown in \cref{fig:llvm-ir}.
The dashed box highlights the instructions to be optimized.
}
\label{subfig:extracted-ir}
\end{subfigure}
\codebox{S2}{E2}
\codebox{S3}{E3}
\begin{subfigure}[b]{\linewidth}
\begin{lstlisting}[language=llvm,numbers=none, xleftmargin=0em,numberblanklines,
    showstringspaces=false, stringstyle=\small\ttfamily, basicstyle=\scriptsize\ttfamily]
define <4 x i8> @src(i64 %a0, ptr %a1) {
entry:
  %0 = getelementptr inbounds nuw i32,ptr %a1,i64 %a0
  %wide.load = load <4 x i32>, ptr %0, align 4
  %smax_0 = smax <4 x i32> %wide.load, zeroinitializer
  %smin_255 = smin <4 x i32> %smax_0, splat (i32 255)
  %result = trunc nuw <4 x i32> %smin_255 to <4 x i8>
  ret <4 x i8> %result
}
\end{lstlisting}
\caption{The syntactically invalid candidate initially output by \geminithinking.}
\label{subfig:incorrect-result}
\end{subfigure}
\end{minipage}
\hfill
\begin{minipage}[b]{0.45\linewidth}
\begin{subfigure}[b]{\linewidth}
\begin{lstlisting}[language=llvm,numbers=none, xleftmargin=0em,numberblanklines,
    showstringspaces=false, stringstyle=\small\ttfamily, basicstyle=\scriptsize\ttfamily]
error: expected instruction opcode
  %smax_0 = smax <4 x i32> %wide.load, zeroinitializer
              ^
\end{lstlisting}
\caption{The error message output by \opt.}
\label{subfig:error-message}
\end{subfigure}

\vspace*{15px}

\begin{subfigure}[b]{\linewidth}
\begin{lstlisting}[language=llvm,numbers=none, xleftmargin=0em,numberblanklines,
    showstringspaces=false, stringstyle=\small\ttfamily, basicstyle=\scriptsize\ttfamily]
define <4 x i8> @src(i64 %a0, ptr %a1) {
entry:
  %0 = getelementptr inbounds nuw i32,ptr %a1,i64 %a0
  %wide.load = load <4 x i32>, ptr %0, align 4
  (*@\tikzmark{S4}@*)%smax_val = tail call <4 x i32> @llvm.smax.v4i32(
    <4 x i32> %wide.load, <4 x i32> zeroinitializer
  )                                                (*@\tikzmark{E4}@*)
  %smin_val = tail call <4 x i32> @llvm.smin.v4i32(
    <4 x i32> %smax_val, <4 x i32> splat (i32 255)
  )
  %result = trunc nuw <4 x i32> %smin_val to <4 x i8>
  ret <4 x i8> %result
}
\end{lstlisting}
\caption{The correct candidate produced by \geminithinking after receiving the error message as feedback.
The dashed box highlights the new instruction that replaces the original
\mycode{icmp} and \mycode{select} instructions.}
\label{subfig:correct-result}
\end{subfigure}
\end{minipage}
\codebox{S4}{E4}
\label{fig:illustrative-example}
\caption{An illustrative example demonstrating the workflow of \proj.}
\end{figure*}

\label{subsec:illustrative:example}
We use the real-world \llvm module
in \cref{fig:llvm-ir} (\cref{sec:background})
to illustrate the workflow of \proj.
Given this module,
\proj first traverses all the basic blocks in each function of this module
and extracts instruction sequences from them.
\cref{subfig:extracted-ir} shows one of the instruction sequences extracted
from the basic block \texttt{vector.body}.

\proj then provides the extracted instruction sequence to an \llm
and
prompts it to produce a correct and optimal instruction sequence.
In this example, \proj uses gemini-2.0-flash-thinking-exp-01-21 (referred to as \geminithinking)
as its underlying \llm,
and the initial candidate produced by this model, shown in \cref{subfig:incorrect-result},
contains a syntax error.

After obtaining this candidate, \proj initiates the verification process.
It first invokes the \llvm optimizer \opt to check candidate's syntax
and to potentially further optimize and canonicalize the candidate.
In this case, since the candidate contains a syntax error, \opt outputs an error message
as shown in
\cref{subfig:error-message}.
\proj then feeds this error message back to the \geminithinking,
prompting it to produce a new candidate.
\cref{subfig:correct-result} shows the corrected candidate generated by \geminithinking after receiving
the feedback.

With this new candidate, \proj again invokes \opt,
and now the syntax check passes.
Meanwhile, no further optimization and canonicalization is performed by \opt since
the candidate is already optimal and canonical.
The subsequent interestingness checking (discussed in \cref{subsec:verification})
also passes, as the candidate
contains one fewer instructions than the original instruction sequence.
Finally, \alive formally confirms the correctness of the proposed optimization,
and \proj records the original instruction sequence and the candidate
as a potential missed optimization.

\souper cannot detect this missed optimization because it does not support the
\llvm intrinsic group \mycode{llvm.umin.*}.

\subsection{Extracting Instruction Sequences}
\label{subsec:extraction}
This subsection introduces the motivation and the design of the extractor  (step \circled{1}).

\myparagraph{Motivation}
The motivation for extracting instruction sequences from an entire \llvm \ir module is threefold.
First, an \llvm \ir module can be extremely large and complex,
and from our experience, existing \llms struggle to identify missed peephole optimizations
directly from a complete module.
By prompting \llms to focus on optimizing small instruction sequences,
we increase the likelihood of discovering potential missed optimizations.
Second, extracting instruction sequences from each IR module allows
\proj to exclude irrelevant information and deduplicate sequences that
appear multiple times in the corpus,
and thereby increase the overall efficiency and save the cost for \llm inference.
Finally, slicing modules into instruction sequences facilitates
subsequent verification and analysis
by restricting the scope to a small instruction sequence in each iteration.

\begin{algorithm}[htbp]
\setstretch{0.85}
\small
\DontPrintSemicolon
\caption{%
    Extracting Instruction Sequences from an \llvm Module -- \protect\AlgExtract{\protect\AlgModule, \protect\AlgDedupSet}}
\label{alg:extractor}
\Input{\AlgModule, an \llvm module}
\Input{\AlgDedupSet, a set that contains all the seen encoded instruction sequences}
\Output{A set of unique instruction sequences wrapped to \llvm IR functions}

\AlgUniqueInstSeqSet $\gets \varnothing$ \; \nllabel{line:init:unique:list}
\ForEach{\AlgFunction $\in$ \AlgModule} { \nllabel{line:traverse:function}
    \ForEach{ basic block \AlgBasicBlock $\in$ \AlgFunction} { \nllabel{line:traverse:basicblock}
    	$\AlgInstSeqList \gets \AlgExtractFromBB{\AlgBasicBlock}$\; \nllabel{line:call:extract}
        \ForEach{\AlgInstSeq $\in$ \AlgInstSeqList} { \nllabel{line:dedup:start}
            \AlgConvertedFunction $\gets$ \AlgConvertToFunction{\AlgInstSeq} \; \nllabel{line:wrap:as:function}
            \If {\AlgConvertedFunction can be further optimized} \Continue \nllabel{line:cannot:further:optimize}
            \AlgEncoding $\gets$ \AlgEncode{\AlgConvertedFunction} \; \nllabel{line:encode}
            \lIf {\AlgEncoding $\in$ \AlgDedupSet} \Continue \nllabel{line:check:duplicate}
            $ \AlgDedupSet \gets \AlgDedupSet \cup \{\AlgEncoding \} $ \;
            $\AlgUniqueInstSeqSet \gets \AlgUniqueInstSeqSet \cup \{ \AlgInstSeq \} $\;
        } \nllabel{line:dedup:end}
    }
}
\Return{\AlgUniqueInstSeqSet} \; \nllabel{line:return}

\BlankLine

\Fn{\AlgExtractFromBB{\AlgBasicBlock}}{
        $\AlgInstSeqList \gets \varnothing$ \; \nllabel{line:init:list}
\ForEach{instruction \AlgInstruction $\in$ \AlgBasicBlock in reverse order} { \nllabel{line:traverse:inst}
	\lIf {\AlgInstruction is a terminator instruction} \Continue \nllabel{line:skip:terminator}
	\AlgUsed $\gets$ \AlgFalse \; \nllabel{line:init:flag}

	$\AlgNewSeqList \gets \varnothing$\;
	\ForEach{instruction sequence $\AlgInstSeq \in \AlgInstSeqList$} { \nllabel{line:traverse:exist:seq}
		\If {any instruction in \AlgInstSeq depends on \AlgInstruction} { \nllabel{line:check:dependency:start}
			$ \AlgNewSeqList \gets \AlgNewSeqList \cup \{ [\AlgInstruction] + \AlgInstSeq \} $\;
			$\AlgUsed \gets \AlgTrue$
		} \nllabel{line:check:dependency:end}
		\lElse{
			$\AlgNewSeqList \gets \AlgNewSeqList \cup \{ \AlgInstSeq \}$ \nllabel{line:check:unchanged}
		}
	}
	\lIf {$\neg$\AlgUsed} { \nllabel{line:if:unused}
		$\AlgNewSeqList \gets \AlgNewSeqList \cup \{ [\AlgInstruction] \}$
	}
	$ \AlgInstSeqList \gets \AlgNewSeqList$\; \nllabel{line:update:set}
}
\Return{\AlgInstSeqList}

}

\end{algorithm}

\cref{alg:extractor} presents the design of \proj's extractor,
whose goal is to obtain all unique dependent instruction
sequences from each basic block in every \llvm \ir module in the corpus.
The extractor takes two inputs:
an \llvm \ir module and a set, \AlgDedupSet
containing all previously seen
encoded instruction sequences for deduplication.
It outputs a set of unique instruction sequences extracted from the given module.
Each of the instruction sequences is wrapped as an \llvm function and passed to the \llm to
explore potential optimization opportunity within it.
The detailed steps are described below.

\myparagraph{Extracting From Basic Blocks}
Given an \llvm IR module,
\proj first initializes \AlgUniqueInstSeqSet to store the extracted sequences (\cref{line:init:unique:list}).
It then traverses each basic block \AlgBasicBlock of each function in the module
(\crefrange{line:traverse:function}{line:traverse:basicblock}),
and extracts all dependent instruction sequences contained within
\AlgBasicBlock by calling function \AlgExtractFromBB (\cref{line:call:extract}).
Specifically, \AlgExtractFromBB
traverses the instructions in \AlgBasicBlock in reverse order
(\cref{line:traverse:inst}).
For each instruction \AlgInstruction,
if it is a terminator (\ie, the last instruction in a basic block),
then it is skipped (\cref{line:skip:terminator}),
since terminators only define control flow jumps,
and \proj currently targets only missed optimization within basic blocks.
On \cref{line:init:flag},
\proj initializes a flag \AlgUsed as \AlgFalse
to indicate whether \AlgInstruction has been inserted into any sequence,
and initialize a new set, \AlgNewSeqList, which is eventually used to update \AlgInstSeq
(\cref{line:update:set}).
Next, for each instruction sequence \AlgInstSeq in \AlgInstSeqList,
if any instruction in \AlgInstSeq depends on \AlgInstruction,
\proj prepends \AlgInstruction to \AlgInstSeq and sets \AlgUsed to \AlgTrue
(\crefrange{line:check:dependency:start}{line:check:dependency:end}).
Otherwise, \AlgInstSeq remains unchanged (\cref{line:check:unchanged}).
If \AlgInstruction is not added to any existing sequences—meaning
no traversed instruction depends on \AlgInstruction—\proj creates
a new sequence containing only \AlgInstruction and adds the new sequence to
\AlgNewSeqList (\cref{line:if:unused}).

\myparagraph{Verifying and Deduplicating Instruction Sequences}
After \proj traverses all the instructions within a basic block,
a set of instruction sequences can be obtained.
For each extracted instruction sequence, \proj wraps it as an \llvm IR function
(\cref{line:wrap:as:function}).
Specifically, \proj appends a return instruction that returns the value
produced by the last instruction to the sequence and
treats all undefined operands in the sequence of instructions
as function arguments.
With the wrapped function obtained, \proj
checks that \llvm cannot further optimize the instruction sequence
(\cref{line:cannot:further:optimize}).
This check is necessary because a sequence that is unoptimizable in its original context
may become optimizable once isolated.
For example, an instruction in the sequence might be used not only by subsequent
instructions within the sequence but also by instructions in later basic blocks;
once those external uses are removed, the instruction could be combined with others in the sequence.
Finally, \proj computes a hash of the function based on the opcode and operands of each instruction
(\cref{line:encode}),
and add instruction sequences whose hashes have not been recorded in \AlgDedupSet
(\crefrange{line:check:duplicate}{line:dedup:end}).

\subsection{Verifying Outputs of the LLM}
\label{subsec:verification}
After extracting instruction sequences from the corpus
and wrapping them as \llvm \ir functions
(we use the term instruction sequence and function interchangeably in the following),
\proj begins prompting the specified \llm to optimize each function.
Each \llm-generated candidate is then verified through three steps.

\myparagraph{Preprocessing with \opt}
First, \proj feeds the candidate to \opt, the \llvm optimizer,
with the \mycode{-O3} flag to perform aggressive optimization (step \circled{3}).
The step serves two purposes.
First, \opt checks whether the function proposed by the \llm is syntactically correct.
If the function contains syntax errors, \opt outputs an error message,
which \proj can then use as feedback to prompt the \llm to fix the error (step \circled{6}).
Second, although the \llm may produce a modified version of the original function,
the proposed function may not be canonical or optimal \wrt \opt.
In such cases, running \opt can further optimize and canonicalize the proposed function
reducing the effort required for subsequent manual analysis.

\myparagraph{Checking Interestingness} \label{subsec:interestingness}
If the function proposed by the \llm is syntactically correct,
\proj next checks whether the function—after preprocessed by
\texttt{opt}—potentially manifests a beneficial optimization.
Determining whether the transformation from
the original function to the proposed one is beneficial is tricky,
because (1) comparison between two functions is not straightforward, as multiple aspects can be
considered (\eg, code size and performance on a specific target);
(2) even if the proposed function appears significantly better,
implementing the corresponding optimization might not be beneficial overall if it
hinders other, more valuable optimizations.
Due to these challenges,
this step is designed to only filter out as many uninteresting
cases as possible
(\ie, those unlikely to manifest a beneficial optimization),
thereby reducing the effort required for subsequent manual analysis.
In our prototype, two metrics are considered when checking the interestingness,
the instruction count and the total cycles measured by
\mca \footnote{A performance analysis tool to statically measure
the performance of machine code in a specific CPU based on \llvm's scheduling models.}
on a specific target with a specific CPU (\eg, with the target triple set to
\mycode{x86\_64-unknown-unknown} and the CPU set to \mycode{btver2})
Functions with fewer instructions or fewer total cycles are
considered interesting and thus pass the check.
Additionally, functions with the same instruction count and total cycles
are also considered interesting if they differ syntactically from the original,
since such transformations—though not immediately improving the function—may enable
further optimization.
Interestingness checking is performed before correctness verification, as it is
generally more efficient, avoiding the %
costlier correctness checking for
unpromising cases and thus improving overall workflow efficiency.

\myparagraph{Verifying Correctness}
If the function proposed by the \llm passes the interestingness checking,
\proj then verifies the correctness of the transformation from the original function
to the proposed one using \alive~\cite{lopes2021alive2}.
If \alive proves the transformation incorrect, it provides a counterexample to demonstrate
the incorrectness, which \proj uses as feedback to prompt the \llm
to propose a new candidate.
If \alive encounters a fixable error that prevents the proof—for example,
mismatched function signatures—the error message is also
used as feedback to guide the \llm in avoiding the issue
(step \circled{6}).
Otherwise, if \alive proves the transformation correct, \proj records the pair of functions
as a potential missed optimization (step \circled{7})

\section{Evaluation}
\label{sec:eval}
To comprehensively evaluate \proj, we conducted experiments to
investigate the following research questions.

\noindent\textbf{RQ1:} Can \proj detect prior missed optimizations?

\noindent\textbf{RQ2:} Can \proj discover new missed optimizations?

\noindent\textbf{RQ3:} How much throughput can \proj achieve, and what is the associated cost?

\subsection{Experimental Setup}

\myparagraph{Selected \llm Models}
\cref{tab:models} lists all the models selected for evaluating \proj.
We selected diverse models to investigate
how the capabilities of different \llms affect the effectiveness of \proj.
For RQ1, we selected two open-source and four proprietary models.
Specifically, we chose two open-source models with significantly different sizes:
Gemma3 (27 billion parameters) and Llama3.3 (70 billion parameters).
For proprietary models,
we included two base models, \ie, \geminitwo and \gptfourone,
and two reasoning models, \ie, \geminithinking and \ofourmini.
These models have relatively early knowledge cut-off dates, allowing us to
collect a reasonable number of benchmarks while avoiding potential data leakage.
In RQ2, during the long-term, intermittent experiment,
we tried several open-source models
and used a locally deployed \vllamathree for most of the time.
In RQ3,
\llamathree and \geminitwofive
(a cost-efficient and low-latency model recommended for scenarios requiring high throughput) are used to evaluate the throughput and cost,
covering two common usage scenarios,
\ie,
using a locally deployed \llm and invoking an \llm via API.
For \geminithinking and \geminitwofive, we set the reasoning budget to \emph{low}
(\ie, at most 1,024 reasoning tokens).

\begin{table}[htbp]
\caption{The selected \llms in evaluation.
\geminitwofive is excluded from RQ1 to prevent potential data leakage.
}
\label{tab:models}
\centering
\small
\renewcommand{\arraystretch}{0.85}
\resizebox{0.47\textwidth}{!}{
\begin{tabular}{@{}llrr@{}}
\toprule
\small{Model Name} & \small{Model Version}           & \small{Reasoning} &  \small{Cut-off Date} \\ \midrule
\gemmathree & \vgemmathree          & No            &  08/2024\\
\llamathree & \vllamathree          & No        & 12/2023\\
\geminitwo & \vgeminitwo         & No        & 08/2024\\
\geminithinking & \begin{tabular}[c]{@{}l@{}}gemini-2.0-flash-\\ thinking-exp-01-21\end{tabular} & Yes       & 08/2024\\
\gptfourone & \vgptfourone & No            & 06/2024\\
\ofourmini & \vofourmini          & Yes       & 06/2024\\ \midrule
\geminitwofive & \vgeminitwofive & Yes      & 01/2025 \\ \bottomrule
\end{tabular}
}

\end{table}
\myparagraph{Baselines} We used three baselines
in our evaluations.

\begin{itemize}[leftmargin=1em, topsep=0pt]
\item \textit{\souper}.
The primary baseline used in our evaluation is
\souper~\cite{sasnauskas2017souper},
the state-of-the-art superoptimizer for \llvm.
\souper allows users to set the maximum instructions for enumerative synthesis
(referred to as \enum).
Increasing this value can improve \souper's capability in finding missed optimization
but also incurs significantly higher computational cost.
We evaluated \souper using different values of \enum:
the default value 0 (\souperdefault) and values from 1 to \souperEnumSynMaxInst (\souperenum).
A timeout of \souperTimeout
is applied to each input to ensure the experiments complete in a reasonable time.

\item \textit{\minotaur}.
We also compared \proj with \minotaur~\cite{liu2024minotaur},
a recent superoptimizer for \llvm that employs program synthesis techniques and
focuses on optimizing integer and floating-point SIMD code.
Default settings were used for \minotaur in our experiments.

\item \textit{\projnofeedback}.
To evaluate the effectiveness of the iterative prompting strategy,
we prepared \projnofeedback, a variant of \proj that does not prompt the \llm to generate new candidates using feedback
(\ie, it omits steps \circled{4} and \circled{7} in \cref{fig:overlall-workflow})

\end{itemize}

\myparagraph{System Configuration}
The experiments were run on an Ubuntu 22.04 server with
three NVIDIA RTX 6000 Ada GPUs,
Intel Xeon Gold 5217 CPU@3.00GHz,
and 384 GB RAM.

\subsection{RQ1: Detecting Previously Reported Missed Optimizations}
\begin{table*}[htbp]

\caption{
Evaluation results on \numberOfBenchmarks previously reported missed optimizations.
For \proj and \projnofeedback, each benchmark is tested with five rounds for each model,
and the results show how many times \proj and \projnofeedback can catch the optimization opportunity.
For \souperenum, if it detects the optimization with \enum set to any value from 1 to 3,
it is marked with a \blackcheck.
Empty cells represent that the optimization is not detected by the tool.
The numbers of successful times are grouped into three categories: \colorbox[HTML]{e5f5e0}{1-2 time},
\colorbox[HTML]{a1d99b}{3-4 times} and all \colorbox[HTML]{31a354}{5 times}.
The \textbf{Average} row shows the average number of successful benchmarks per round for \proj and \projnofeedback,
and the \textbf{Total} row shows the total number of missed optimizations that are detected at least once.
}
\small
\renewcommand{\arraystretch}{0.85}
\resizebox{\textwidth}{!}{
\begin{tabular}{@{}l|rr|rr|rr|rr|rr|rr|cc|c@{}}
\toprule
\multirow{2}{*}{Issue ID} & \multicolumn{2}{|c}{Gemma3} & \multicolumn{2}{|c}{Llama3.3} & \multicolumn{2}{|c}{Gemini2.0} & \multicolumn{2}{|c}{Gemini2.0T} & \multicolumn{2}{|c}{GPT-4.1} & \multicolumn{2}{|c}{o4-mini} & \multicolumn{2}{|c}{Souper} & \multicolumn{1}{|c}{Minotaur} \\ \cmidrule(lr){2-16}
        & \multicolumn{1}{|c}{\projnofeedback} & \multicolumn{1}{c}{\proj} & \multicolumn{1}{|c}{\projnofeedback} & \multicolumn{1}{c}{\proj} & \multicolumn{1}{|c}{\projnofeedback} & \multicolumn{1}{c}{\proj} & \multicolumn{1}{|c}{\projnofeedback} & \multicolumn{1}{c}{\proj} & \multicolumn{1}{|c}{\projnofeedback} & \multicolumn{1}{c}{\proj} & \multicolumn{1}{|c}{\projnofeedback} & \multicolumn{1}{c}{\proj} & \multicolumn{1}{|c}{\default} & \multicolumn{1}{c}{\enum} & \multicolumn{1}{|c}{\default} \\ \midrule
\issueId{104875} &                           &                              &                              &                             &                            &                            & \cellcolor[HTML]{e5f5e0}1    & \cellcolor[HTML]{31a354}5  &                            &                              &                            &                             &             &             &             \\
\issueId{107228} &                           &                              &                              &                             &                            &                            & \cellcolor[HTML]{31a354}5    & \cellcolor[HTML]{31a354}5  &                            & \cellcolor[HTML]{e5f5e0}1    & \cellcolor[HTML]{a1d99b}4  & \cellcolor[HTML]{31a354}5   &             & \blackcheck &             \\ %
\issueId{108451} &                           &                              & \cellcolor[HTML]{31a354} 5   & \cellcolor[HTML]{31a354}5   & \cellcolor[HTML]{31a354}5  & \cellcolor[HTML]{31a354}5  & \cellcolor[HTML]{a1d99b}4    & \cellcolor[HTML]{31a354}5  & \cellcolor[HTML]{e5f5e0}1  & \cellcolor[HTML]{a1d99b}4    & \cellcolor[HTML]{e5f5e0}2  & \cellcolor[HTML]{31a354}5   &             & \blackcheck & \blackcheck \\
\issueId{108559} &                           &                              & \cellcolor[HTML]{31a354} 5   & \cellcolor[HTML]{31a354}5   & \cellcolor[HTML]{31a354}5  & \cellcolor[HTML]{31a354}5  & \cellcolor[HTML]{a1d99b}4    & \cellcolor[HTML]{31a354}5  & \cellcolor[HTML]{e5f5e0}1  & \cellcolor[HTML]{a1d99b}4    & \cellcolor[HTML]{a1d99b}4  & \cellcolor[HTML]{31a354}5   &             & \blackcheck &             \\
\issueId{110591} &                           &                              & \cellcolor[HTML]{31a354} 5   & \cellcolor[HTML]{31a354}5   & \cellcolor[HTML]{31a354}5  & \cellcolor[HTML]{31a354}5  & \cellcolor[HTML]{e5f5e0}2    & \cellcolor[HTML]{31a354}5  & \cellcolor[HTML]{e5f5e0}2  & \cellcolor[HTML]{31a354}5    & \cellcolor[HTML]{a1d99b}3  & \cellcolor[HTML]{31a354}5   &             &             &             \\ %
\issueId{115466} & \cellcolor[HTML]{e5f5e0}1 & \cellcolor[HTML]{e5f5e0}1    & \cellcolor[HTML]{31a354} 5   & \cellcolor[HTML]{31a354}5   &                            &                            & \cellcolor[HTML]{31a354}5    & \cellcolor[HTML]{31a354}5  & \cellcolor[HTML]{a1d99b}3  & \cellcolor[HTML]{a1d99b}4    & \cellcolor[HTML]{31a354}5  & \cellcolor[HTML]{31a354}5   &             & \blackcheck &             \\
\issueId{118155} &                           &                              &                              &                             &                            &                            & \cellcolor[HTML]{a1d99b}3    & \cellcolor[HTML]{a1d99b}3  &                            &                              & \cellcolor[HTML]{e5f5e0}2  & \cellcolor[HTML]{e5f5e0}2   &             &             &             \\
\issueId{122235} &                           &                              &                              & \cellcolor[HTML]{31a354}5   & \cellcolor[HTML]{e5f5e0}2  & \cellcolor[HTML]{31a354}5  &                              & \cellcolor[HTML]{e5f5e0}2  &                            & \cellcolor[HTML]{31a354}5    &                            &                             &             & \blackcheck &             \\
\issueId{122388} &                           &                              &                              &                             & \cellcolor[HTML]{31a354}5  & \cellcolor[HTML]{31a354}5  & \cellcolor[HTML]{a1d99b}4    & \cellcolor[HTML]{a1d99b}4  & \cellcolor[HTML]{e5f5e0}2  & \cellcolor[HTML]{a1d99b}3    &                            &                             &             &             &             \\
\issueId{126056} &                           &                              &                              &                             &                            & \cellcolor[HTML]{e5f5e0}1  & \cellcolor[HTML]{31a354}5    & \cellcolor[HTML]{31a354}5  & \cellcolor[HTML]{e5f5e0}1  & \cellcolor[HTML]{a1d99b}4    & \cellcolor[HTML]{31a354}5  & \cellcolor[HTML]{31a354}5   &             & \blackcheck & \blackcheck \\
\issueId{128475} &                           &                              &                              &                             &                            &                            & \cellcolor[HTML]{e5f5e0}2    & \cellcolor[HTML]{e5f5e0}2  &                            &                              & \cellcolor[HTML]{a1d99b}4  & \cellcolor[HTML]{31a354}5   &             & \blackcheck &             \\
\issueId{128778} &                           &                              &                              &                             &                            & \cellcolor[HTML]{e5f5e0}1  &                              & \cellcolor[HTML]{31a354}5  &                            & \cellcolor[HTML]{e5f5e0}1    & \cellcolor[HTML]{a1d99b}3  & \cellcolor[HTML]{31a354}5   &             & \blackcheck &             \\
\issueId{129947} &                           &                              &                              &                             &                            &                            &                              & \cellcolor[HTML]{e5f5e0}1  &                            &                              &                            &                             &             &             &             \\
\issueId{131444} &                           &                              &                              &                             &                            &                            &                              &                            &                            &                              &                            &                             &             &             &             \\
\issueId{131824} &                           &                              &                              &                             &                            & \cellcolor[HTML]{e5f5e0}1  &                              & \cellcolor[HTML]{a1d99b}3  &                            & \cellcolor[HTML]{e5f5e0}1    &                            & \cellcolor[HTML]{a1d99b}3   &             & \blackcheck &             \\
\issueId{132508} &                           & \cellcolor[HTML]{e5f5e0}2    & \cellcolor[HTML]{e5f5e0}1    & \cellcolor[HTML]{31a354}5   & \cellcolor[HTML]{31a354}5  & \cellcolor[HTML]{31a354}5  & \cellcolor[HTML]{e5f5e0}2    & \cellcolor[HTML]{e5f5e0}2  &                            & \cellcolor[HTML]{a1d99b}3    & \cellcolor[HTML]{a1d99b}3  & \cellcolor[HTML]{31a354}5   & \blackcheck &             &             \\
\issueId{134318} &                           &                              &                              &                             &                            &                            &                              &                            &                            &                              &                            &                             &             &             &             \\
\issueId{135411} &                           &                              &                              &                             &                            &                            & \cellcolor[HTML]{31a354}5    & \cellcolor[HTML]{31a354}5  & \cellcolor[HTML]{e5f5e0}1  & \cellcolor[HTML]{e5f5e0}1    & \cellcolor[HTML]{31a354}5  & \cellcolor[HTML]{31a354}5   &             & \blackcheck & \blackcheck \\
\issueId{137161} &                           &                              &                              &                             &                            & \cellcolor[HTML]{e5f5e0}2  &                              &                            &                            &                              &                            & \cellcolor[HTML]{e5f5e0}1   &             &             &             \\
\issueId{141479} &                           &                              &                              &                             &                            &                            & \cellcolor[HTML]{31a354}5    & \cellcolor[HTML]{31a354}5  &                            &                              & \cellcolor[HTML]{a1d99b}4  & \cellcolor[HTML]{31a354}5   & \blackcheck & \blackcheck &             \\
\issueId{141753} &                           &                              &                              &                             &                            &                            &                              & \cellcolor[HTML]{e5f5e0}1  &                            &                              & \cellcolor[HTML]{e5f5e0}1  & \cellcolor[HTML]{e5f5e0}2   &             & \blackcheck &             \\
\issueId{141930} & \cellcolor[HTML]{a1d99b}3 & \cellcolor[HTML]{a1d99b}3    & \cellcolor[HTML]{a1d99b}4    & \cellcolor[HTML]{31a354}5   & \cellcolor[HTML]{e5f5e0}2  & \cellcolor[HTML]{e5f5e0}2  & \cellcolor[HTML]{31a354}5    & \cellcolor[HTML]{31a354}5  &                            &                              & \cellcolor[HTML]{31a354}5  & \cellcolor[HTML]{31a354}5   &             & \blackcheck &             \\
\issueId{142497} &                           &                              &                              &                             &                            &                            &                              & \cellcolor[HTML]{e5f5e0}1  &                            & \cellcolor[HTML]{e5f5e0}1    &                            & \cellcolor[HTML]{e5f5e0}1   &             &             &             \\
\issueId{142593} &                           &                              &                              &                             &                            &                            &                              & \cellcolor[HTML]{a1d99b}4  &                            &                              &                            & \cellcolor[HTML]{e5f5e0}2   & \blackcheck & \blackcheck &             \\
\issueId{143259} &                           &                              &                              &                             &                            &                            &                              &                            &                            &                              &                            &                             &             &             &             \\ \midrule
Average & 0.8 & 1.2 & 5.2 & 7 & 5.8 & 7.4 & 10.4 & 15.6 & 2.2 & 7.4 & 10.0 & 14.2 & N/A & N/A & N/A \\ \midrule
Total  & 2 & 3 & 6 & 7 & 7 & 11 & 14 & 21 & 7 & 12 & 14 & 18 & 3 & 14 & 3 \\ \bottomrule
\end{tabular}
}
\label{tab:rq1}
\end{table*}
To quantitatively evaluate the effectiveness of \proj in finding missed optimization opportunities
and to objectively compare it with the baselines,
we conducted a controlled experiment to assess how effectively each tool identifies
previously reported missed optimizations.
For the benchmark suite,
we collected \numberOfBenchmarks missed peephole optimizations
from issues in the official \llvm GitHub repository tagged with
both \emph{missed-optimization} and \emph{llvm:instcombine} labels.
To avoid potential data leakage~\cite{DBLP:journals/corr/abs-2502-06215, DBLP:journals/tse/LopezCSSV25},
we only included issues created after the end of August 2024,
which is the latest knowledge cut-off date among all \llms listed in \cref{tab:models}
except for \geminitwofive, the model excluded from this research question.
For each benchmark, we collected the example suboptimal \llvm \ir function provided in the issue
and examined whether \proj, \projnofeedback, \souper and \minotaur
could detect the
missed optimization within the function.
To mitigate the effects of the inherent non-determinism of \llms,
we repeated each experiment five times for both \proj and \projnofeedback.
The results of the experiment are shown in \cref{tab:rq1}.

\myparagraph{Effectiveness of \proj with Different \llms}
The effectiveness of \proj heavily depends on the code understanding and reasoning capabilities
of the employed \llm, and \llms with different capabilities can therefore lead to
varying levels of effectiveness.
As shown in \cref{tab:rq1}, the performance of \proj varies significantly across different \llms.
With \gemmathree, the smallest and weakest of the selected models,
\proj can detect at most 3 out of \numberOfBenchmarks missed optimizations,
and on average only 1.2 per round.
With other non-reasoning but more capable models,
the performance of \proj improves substantially:
using \llamathree, \geminitwo, and \gptfourone,
\proj can identify up to 7, 11 and 12 missed optimizations, and on average 7.0, 7.4, and 7.4, respectively.
Employing reasoning models further enhance the performance of \proj.
With \geminithinking and \ofourmini, \proj can
detect up to 21 and 18 missed optimizations,
and on average 15.6 and 14.2, respectively.

\myparagraph{\proj \vs \souper and \minotaur}
We evaluated \souper with maximum instructions
for enumerative synthesis set to 0 (\souperdefault) and from 1 to
\souperEnumSynMaxInst (\souperenum).
\souperdefault, designed for efficiency,
can identify only 3 out of \numberOfBenchmarks missed optimizations,
while \souperenum, with different values of \enum,
detected up to 14 out of \numberOfBenchmarks at the cost of
longer execution time.
In total, \souper detected 15 missed optimizations
(for Issue 132508,
\souperenum either cannot detect or crashed but
\souperdefault can successfully handle).
The most common reason that \souper fails to detect a missed optimization is that
the instruction sequence contains instructions unsupported by \souper.
Among the \numberOfBenchmarksFormalCannotDetect missed optimizations that \souper cannot detect,
\proj can detect up to \numberOfBenchmarksSouperCannotLLMsCan of them.
In this benchmark, \minotaur detects 3 missed optimizations, all of
which are also identified by \souperenum and by \proj with a reasoning model.
These results suggest that, despite its optimization for integer and floating-point SIMD code,
\minotaur remains limited in detecting missed peephole optimizations.

\myparagraph{\proj \vs \projnofeedback}
The iterative prompting strategy improves \proj's effectiveness
in identifying missed optimizations regardless of the model employed.
Without iterative prompting, \projnofeedback
failed to identify 1 to 7 optimization that \proj can detect across different \llms.
Moreover, the average number of optimizations detected by \projnofeedback
is consistently lower than that of \proj, with the gaps ranging from 0.4 to 5.2.

\begin{table*}[htbp]
\caption{All missed optimizations found by \proj and reported to \llvm.
The columns \textbf{\souperdefault}, \textbf{\souperenum}, and \textbf{\minotaur} indicate
whether each tool can detect the corresponding missed optimization.
The result ``timeout'' means \souperenum cannot detect the missed optimization within a \souperTimeoutAdj timeout with \enum set to any value from 1 to 3.
Out of 62 issues in total, 28 are confirmed, 13 have been fixed, 4 are duplicates, and 3 are marked as ``wontfix''.
}
\label{tab:reported}
\small
\renewcommand{\arraystretch}{0.85}
\begin{minipage}[t]{0.48\textwidth}
\centering
\resizebox{\linewidth}{!}{%
\begin{tabular}{lrccc}
\toprule
Issue ID & Status & \souperdefault & \souperenum & \minotaur \\ \midrule
\issueId{128134}   & Fixed       &     &    & \\
\issueId{128460}   & Confirmed   &     & timeout   & \\
\issueId{130954}   & Wontfix     &     &    & \\
\issueId{132628}   & Wontfix     &     & timeout   & \\
\issueId{133367}   & Fixed       &     &    & \\
\issueId{139641}   & Confirmed   &     & \blackcheck & \blackcheck \\
\issueId{139786}   & Confirmed   &     &    & \\
\issueId{142674}   & Fixed       & \blackcheck  & \blackcheck & \\
\issueId{142711}   & Fixed       &     &    & \\
\issueId{143030}   & Unconfirmed &     &    & \\
\issueId{143211}   & Fixed       &     & \blackcheck & \\
\issueId{143630}   & Unconfirmed &     & \blackcheck & \\
\issueId{143636}   & Fixed       &     & \blackcheck & \\
\issueId{143649}   & Unconfirmed &     &    & \\
\issueId{143957}   & Confirmed   & \blackcheck  & timeout   & \\
\issueId{144020}   & Confirmed   &     & \blackcheck & \blackcheck \\
\issueId{152237}   & Confirmed   &     & \blackcheck & \blackcheck \\
\issueId{152788}   & Unconfirmed &     &    & \blackcheck \\
\issueId{152797}   & Confirmed   &     & timeout   & \\
\issueId{152804}   & Confirmed   & \blackcheck  & \blackcheck & \blackcheck \\
\issueId{153991}   & Confirmed   &     &    & \\
\issueId{153999}   & Duplicate   &     &    & \\ %
\issueId{154000}   & Duplicate   & \blackcheck & \blackcheck & \\ %
\issueId{154025}   & Unconfirmed & \blackcheck &  & \\
\issueId{154035}   & Unconfirmed &     &    & \\
\issueId{154238}   & Fixed       &     &    & \\ %
\issueId{154242}   & Confirmed   &     &    & \blackcheck \\
\issueId{154246}   & Confirmed   &     &    & \\
\issueId{154258}   & Unconfirmed & \blackcheck    &  \blackcheck  & \\
\issueId{157315}   & Fixed       &     &  & \\
\issueId{157370}   & Fixed       &     &  \blackcheck  & \\
\bottomrule
\end{tabular}
}
\end{minipage}\hfill
\begin{minipage}[t]{0.48\textwidth}
\centering
\resizebox{\linewidth}{!}{%
\begin{tabular}{lrccc}
\toprule
Issue ID & Status & \souperdefault & \souperenum & \minotaur \\ \midrule
\issueId{157371}   & Fixed       &     &  & \\
\issueId{157372}   & Duplicate   &     &  & \\
\issueId{157486}   & Confirmed   &     &  & \\
\issueId{157524}   & Fixed       &     &  & \\
\issueId{163084}   & Confirmed   &     & \blackcheck & \\
\issueId{163093}   & Unconfirmed &     &  & \\
\issueId{163108}   & Fixed       &     &  \blackcheck & \blackcheck \\
\issueId{163109}   & Confirmed   &     &  & \\
\issueId{163110}   & Confirmed   &     & \blackcheck  & \\
\issueId{163112}   & Confirmed   &     &  & \\
\issueId{163115}   & Confirmed   &     &  & \\
\issueId{166878}   & Confirmed   &     &  & \\
\issueId{166885}   & Confirmed   &     &  & \\
\issueId{166887}   & Unconfirmed   &     & \blackcheck & \blackcheck \\
\issueId{166890}   & Unconfirmed   &     & \blackcheck & \\
\issueId{166973}   & Fixed       &     &  & \blackcheck \\
\issueId{167003}   & Confirmed   &     &  & \blackcheck \\
\issueId{167014}   & Confirmed   &     &  & \\
\issueId{167055}   & Confirmed   &     &  & \\
\issueId{167059}   & Unconfirmed &     &  & \\
\issueId{167079}   & Unconfirmed &     &  & \\
\issueId{167090}   & Unconfirmed   &     &  & \blackcheck \\
\issueId{167094}   & Duplicate   &     &  & \\
\issueId{167096}   & Confirmed   &     &  & \\
\issueId{167173}   & Confirmed   &     & \blackcheck & \blackcheck \\
\issueId{167178}   & Unconfirmed   &     & \blackcheck & \\
\issueId{167183}   & Confirmed   &     & \blackcheck & \blackcheck \\
\issueId{167190}   & Confirmed   &     &  & \\
\issueId{167199}   & Wontfix     &     &  & \\
\issueId{170020}   & Confirmed   &     & \blackcheck & \\
\issueId{170071}   & Confirmed   &     &  & \\
\bottomrule
\end{tabular}
}
\end{minipage}
\end{table*}

\subsection{RQ2: Discovering New Missed Optimization}
To demonstrate the practical value of \proj, we intermittently ran
an experiment that employs \proj to search for new missed optimizations in \llvm.
In this experiment,
the corpus used by \proj is from an \llvm IR dataset named \optbenchmark~\cite{llvm-opt-benchmark}.
This dataset contains optimized IR files from \optbenchmarkNumProjects real world projects written in
C, C++, or Rust.
Since the entire benchmark is too massive to serve as our corpus,
we built our corpus by selecting five popular projects for each programming language,
\ie, \mycode{cpython}, \mycode{ffmpeg}, \mycode{linux}, \mycode{openssl}, \mycode{redis},
\mycode{node}, \mycode{protobuf}, \mycode{opencv}, \mycode{z3}, \mycode{pingora},
\mycode{ripgrep}, \mycode{typst}, \mycode{uv}, \mycode{zed}.
From this corpus, \proj extracted over \numberOfExtractedSnippets unique instruction sequences,
eliminating around \numberOfDuplicates duplicates.

\cref{tab:reported} lists all missed optimizations found by \proj and reported to \llvm.
Within \monthsOfExperiment months of development and intermittent testing, \proj has
found \reportedInTotal potential missed peephole optimizations,
with \confirmedBugs confirmed and \fixedBugs already fixed in \llvm.
It also detected a bug in \alive.\footnote{\alivebuglink}
Among all the reported missed optimizations,
we marked \wontFix of them as \emph{wontfix} (\ie, \wontfixone, \wontfixtwo, and \wontfixthree)
according to the feedback from the maintainers.
The reasons are: (1) the optimization is handled by the backend,
(2) the optimization would prevent other valuable optimizations,
and (3) the optimization is application-specific and is not general enough.

We also investigated if \souper and \minotaur can detect these reported missed optimizations.
As shown in \cref{tab:reported}, \souperdefault can identify \reportedDefaultCanFind
out of \reportedInTotal missed optimization, \confirmedDefaultCanFind of which are confirmed or fixed;
\souperenum can identify \reportedEnumCanFind, and \confirmedEnumCanFind of them are confirmed or fixed;
\minotaur can identify \reportedMinotaurCanFind, and \confirmedMinotaurCanFind of them are confirmed or fixed.
In the remainder of the section, we present three examples of
confirmed missed optimizations found by \proj
that neither \souper nor \minotaur can detect.

\myparagraph{Case Study 1 (\cref{subfig:casestudy1:before,subfig:casestudy1:after})}
The \mycode{src} function loads two consecutive 16-bit values
from a given address \mycode{\%0}
and combines them into a single 32-bit value using zero extension (\mycode{zext}),
left shift (\mycode{shl}), and an \mycode{or} instruction.
This function can be optimized to directly loading a 32-bit value from the given address.
\llvm originally missed this optimization because it did not support merging consecutive loads with different sizes.
Specifically, the original unoptimized function
loads three consecutive values—two 8-bit values and one 16-bit value.
\llvm successfully merges the two 8-bit loads but fails to further merge them with
the subsequent 16-bit load.
\souper and \minotaur cannot identify this case, as they cannot synthesize \mycode{load}
and \mycode{getelementptr} instructions.

\myparagraph{Case Study 2 (\cref{subfig:casestudy2:before,subfig:casestudy2:after})}
In the \mycode{src} function, the input \mycode{\%0} is first clamped to be at least \mycode{1}
by \mycode{llvm.umax.i8},
then doubled with a 1-bit left shift,
and finally clamped again to be at least \mycode{16}.
In the \mycode{tgt} function, the initial clamp is removed.
This simplification is valid because the second clamp to \mycode{16} subsumes the effect of the first clamp,
which makes it redundant.
\souper cannot identify this missed optimization because it does not support the intrinsic family \mycode{llvm.umax.*}.
Although \minotaur supports synthesizing this operation, it fails to detect the missed optimization.

\myparagraph{Case Study 3 (\cref{subfig:casestudy3:before,subfig:casestudy3:after})}
In the \mycode{src} function, the input \mycode{\%0} is first checked for being an ordinary (non-NaN) value;
if it is NaN, it is replaced with \mycode{0.0}, otherwise, \mycode{\%0} is kept.
The result is then compared against \mycode{1.0}.
In the \mycode{tgt} function, the check and replacement are removed,
and \mycode{\%0} is directly compared with \mycode{1.0}.
This simplification is valid because the instruction \mycode{fcmp oeq} (ordered equal)
already returns false if \mycode{\%0} is NaN,
so explicitly substituting NaNs with \mycode{0.0} is redundant.
\souper cannot identify this missed optimization because it does not support floating-point instructions,
and \minotaur crashes on this IR function.

\begin{figure*}[htbp]
\centering
\begin{tabular}{@{}c@{}|@{}c@{}|@{}c@{}}
\begin{subfigure}[b]{0.32\linewidth}
\begin{lstlisting}[language=llvm,numbers=none, xleftmargin=0em,numberblanklines,
    showstringspaces=false, stringstyle=\small\ttfamily, basicstyle=\scriptsize\ttfamily]
define i32 @src(ptr %0) {
  %2 = load i16, ptr %0, align 2
  %3 = getelementptr i8, ptr %0, i64 2
  %4 = load i16, ptr %3, align 1
  %5 = zext i16 %4 to i32
  %6 = shl nuw i32 %5, 16
  %7 = zext i16 %2 to i32
  %8 = or disjoint i32 %6, %7
  ret i32 %8
}
\end{lstlisting}
\caption{Case 1: sub-optimal version.}
\label{subfig:casestudy1:before}
\end{subfigure}
&
\begin{subfigure}[b]{0.3\linewidth}
\begin{lstlisting}[language=llvm,numbers=none, xleftmargin=1em,numberblanklines,
    showstringspaces=false, stringstyle=\small\ttfamily, basicstyle=\scriptsize\ttfamily]
define i8 @src(i8 %0) {
  %2 = call i8 @llvm.umax.i8(
    i8 %0, i8 1)
  %3 = shl nuw i8 %2, 1
  %4 = call i8 @llvm.umax.i8(
    i8 %3, i8 16)
  ret i8 %4
}
\end{lstlisting}
\caption{Case 2: sub-optimal version.}
\label{subfig:casestudy2:before}
\end{subfigure}
&
\begin{subfigure}[b]{0.35\linewidth}
\begin{lstlisting}[language=llvm,numbers=none, xleftmargin=1em,numberblanklines,
    showstringspaces=false, stringstyle=\small\ttfamily, basicstyle=\scriptsize\ttfamily]
define i1 @src(double %0) {
  %2 = fcmp ord double %0, 0.000000e+00
  %3 = select i1 %2,
        double %0, double 0.000000e+00
  %4 = fcmp oeq double %3, 1.000000e+00
  ret i1 %4
}
\end{lstlisting}
\caption{Case 3: sub-optimal version.}
\label{subfig:casestudy3:before}
\end{subfigure}
\\
\begin{subfigure}[b]{0.32\linewidth}
\begin{lstlisting}[language=llvm,numbers=none, xleftmargin=0em,numberblanklines,
    showstringspaces=false, stringstyle=\small\ttfamily, basicstyle=\scriptsize\ttfamily]
define i32 @tgt(ptr %0) {
  %2 = load i32, ptr %0, align 2
  ret i32 %2
}
\end{lstlisting}
\caption{Case 1: optimized version.}
\label{subfig:casestudy1:after}
\end{subfigure}
&
\begin{subfigure}[b]{0.3\linewidth}
\begin{lstlisting}[language=llvm,numbers=none, xleftmargin=1em,numberblanklines,
    showstringspaces=false, stringstyle=\small\ttfamily, basicstyle=\scriptsize\ttfamily]
define i8 @tgt(i8 %0) {
  %2 = shl nuw i8 %0, 1
  %3 = call i8 @llvm.umax.i8(
    i8 %2, i8 16)
  ret i8 %3
}
\end{lstlisting}
\caption{Case 2: optimized version.}
\label{subfig:casestudy2:after}
\end{subfigure}
&
\begin{subfigure}[b]{0.35\linewidth}
\begin{lstlisting}[language=llvm,numbers=none, xleftmargin=1em,numberblanklines,
    showstringspaces=false, stringstyle=\small\ttfamily, basicstyle=\scriptsize\ttfamily]
define i1 @tgt(double %0) {
  %2 = fcmp oeq double %0, 1.000000e+00
  ret i1 %2
}
\end{lstlisting}
\caption{Case 3: optimized version.}
\label{subfig:casestudy3:after}
\end{subfigure}
\end{tabular}
\label{fig:casestudy}
\caption{Three examples of confirmed missed optimizations found by \proj that \souper and \minotaur fail to detect.}
\end{figure*}

\begin{table}[]
\caption{Average case execution time (including timeouts) of \proj (with \llamathree and \geminitwofive) \vs
\souperdefault and \souperenum.
We set the timeout to \souperTimeout,
and the number of timeouts for each tool is also listed.
}
\renewcommand{\arraystretch}{0.85}
\resizebox{\linewidth}{!}{
\begin{tabular}{@{}c|cc|cccc@{}}
\toprule
\multirow{2}{*}{Tool} & \multicolumn{2}{c|}{\proj}                          & \multicolumn{4}{c}{\souper}                      \\ \cmidrule(l){2-7}
                      & \llamathree & \geminitwofive & \default & \enum=1 & \enum=2 & \enum=3 \\ \midrule
Time/Case (s)         & \timeOfLPOllama & \timeOfLPOgemini & \timeOfSouperdefault & \timeOfSouperenumOne & \timeOfSouperenumTwo & \timeOfSouperenumThree\\ \midrule
\# of Timeouts  & 0 & 0 & 0 & \numberOfSouperenumoneTimeout &  \numberOfSouperenumtwoTimeout & \numberOfSouperenumthreeTimeout\\ \bottomrule
\end{tabular}
}
\label{tab:throughput}
\end{table}
\subsection{RQ3: Throughput and Cost of \proj}
To evaluate the throughput and cost of \proj, we constructed a benchmark suite consisting of
\numberOfThroughputBenchmarks instruction sequences randomly sampled from those extracted from
aforementioned real-world projects.
We measured the throughput of \proj under two different setups—using a locally deployed \llamathree and
using \geminitwofive via API.
For the latter setup, we also recorded the API cost.
For comparison,
we evaluated \souperdefault and \souperenum
on the same benchmark suite as baselines.
All tools were executed with a single thread
and the results are presented in \cref{tab:throughput}.

Overall, the throughput of \proj under both setups was lower than
\souperdefault but higher than \souperenumone.
Specifically, \proj with a locally deployed \llamathree took
\totalTimeOfLPOllama in total to process all the cases,
averaging \timeOfLPOllama seconds per case.
In contrast,
\proj with \geminitwofive via API achieved a significantly higher throughput,
completing the experiment in
\totalTimeOfLPOgemini with an average of \timeOfLPOgemini seconds per case,
at a total cost of approximately \costOfLPOgemini.

For \souper, the default configuration achieved the highest throughput,
processing all cases in about
\totalTimeOfSouperdefault, averaging \timeOfSouperdefault seconds per case.
However, as the \enum parameter increases, the throughput decreases sharply.
With \enum set to 1, 2, and 3, the execution time rose to
\totalTimeOfSouperenumone, \totalTimeOfSouperenumtwo, and \totalTimeOfSouperenumthree,
averaging \timeOfSouperenumOne, \timeOfSouperenumTwo, and \timeOfSouperenumThree, respectively.
Moreover, \souper failed to terminate within \souperTimeout in \numberOfSouperenumoneTimeout,
\numberOfSouperenumtwoTimeout, and \numberOfSouperenumthreeTimeout cases
for \enum value of 1, 2, and 3, respectively.

In summary, these results further demonstrate the feasibility
and practical value of \proj.
Although \proj requires additional computation resources for \llm inference compared to \souper,
cost-efficient commercial \llms (\eg, \geminitwofive from Google) enables \proj to achieve
high throughput with manageable monetary cost.
Moreover, as \llms continue to become faster and more cost-effective,
the overall cost-effectiveness of \proj will likewise improve.

\subsection{Threats to Validity}
The first threat to validity is potential data leakage.
When evaluating the effectiveness of \proj in finding previously reported
missed optimization, it is possible that the training set of the deployed model
already contains information related to the missed optimizations,
which could inflate the results. To mitigate this threat, we selected \llms with clear
knowledge cut-off dates and only included missed optimizations reported after those
dates in our benchmarks.
Another concern is that the collected previously reported missed optimizations may not
be fully representative. To mitigate this, we gathered as many benchmarks as possible and
manually inspected each case to ensure validity.
The non-determinism of \llms poses a threat to the internal validity.
We mitigate this by evaluating \proj with multiple \llms and repeating
the controlled experiment in RQ1 for five times.

\section{Discussion}
\label{sec:discussion}
In this section, we discuss the impact of the optimizations
discovered by \proj, the key conceptual contribution of the work,
and the future work on this research direction.

\subsection{Impact of the Discovered Optimizations}
In LLVM development practice, each patch implementing
a peephole optimization is typically evaluated
using \optbenchmark~\cite{llvm-opt-benchmark}, a dataset of \llvm \ir files
collected from \optbenchmarkNumProjects real-world projects written in C, C++ or Rust, and
the \llvm compile time tracker~\cite{llvm-compile-time-tracker}.
Maintainers carefully review each patch and accept it only if:
\circled{1} the net effect is positive, as demonstrated in the benchmark; and
\circled{2} the compile-time impact is acceptable.
In other words, each fixed missed optimization demonstrates tangible benefits in practice anticipated by the community.
To date, \fixedBugs reported missed optimizations have been fixed with 15 carefully reviewed patch commits,
demonstrating that optimizations
discovered by LPO have a real-world impact.

\begin{table}[!ht]
    \caption{The Number of impacted IR files and projects and compile time impact for each
    accepted patch.
    }
    \small
    \centering
    \renewcommand{\arraystretch}{0.85}
    \begin{tabular}{lll c}
    \toprule
    \multirow{2}{*}{ID} & \multirow{2}{*}{\#IR Files} & \multirow{2}{*}{\#Projects} & {$\Delta$ in Compile Time} \\
    & & &(instruction:u) \\
    \midrule
    \issueId{128134} & 54 & 13 & +0.02\% \\
    \issueId{133367} & 122 & 18 & N/A \\
    \issueId{142674} & 251 & 15 & +0.05\% \\
    \issueId{142711} & 10 & 1 & -0.00\% \\
    \issueId{143211} & 16 & 4 & N/A \\
    \issueId{143636} & 2,476 & 68 & +0.02\% \\
    \issueId{154238} & 10 & 4 & N/A \\
    \issueId{157315} & 6 & 2 & +0.00\% \\
    \issueId{157370} & N/A & N/A & +0.04\% \\
    \issueId{157371} (1) & 10 & 13 & N/A \\
    \issueId{157371} (2) & 28 & 1 & +0.02\% \\
    \issueId{157524} & N/A & N/A & -0.03\% \\
    \issueId{163108} (1) & 3,055 & 93 & -0.05\% \\
    \issueId{163108} (2) & 28 & 4 & -0.01\% \\
    \issueId{166973} & 759 & 62 & N/A \\
    \bottomrule
    \end{tabular}
    \label{tab:impacted-ir-files}
\end{table}
\myparagraph{Prevalence of the Detected Suboptimal Patterns}
We analyzed the evaluation results of the patches that fixed the \fixedBugs fixed missed optimization
using \optbenchmark,
recording the number of impacted IR files and projects to
estimate the prevalence of the corresponding suboptimal patterns in real-world code.
As shown in \cref{tab:impacted-ir-files},
most of these handled suboptimal patterns appear in multiple real-world projects.
Notably, Issue \anonymizedIssueId{143636} affects 2,476 IR files across 68 projects,
and Issue \anonymizedIssueId{163108} affects over 3,000 files across 97 projects.
These results highlight that LPO can identify impactful missed optimizations.

\myparagraph{Impacts on the Compile Time}
For each patch, we also examined the compile-time impact using the \llvm compiler time tracker.
\cref{tab:impacted-ir-files} reports the geometric mean of the compile-time changes
across 10 realistic and representative benchmarks
(kimwitu++, sqlite3, consumer-typeset, Bullet, tramp3d-v4, mafft, ClamAV, lencod, SPASS, and 7zip)
comparing the compiler before and after the patch.
The compile-time metric is measured as \texttt{instruction:u},
i.e., the number of machine instructions executed in user space during compilation.
A positive change indicates a slowdown.
As the table shows, the compile-time impact of each patch is negligible.

\begin{figure}[!ht]
    \centering
\ifx\useAnonymizedIssueId\undefined%
    \includegraphics[width=\columnwidth]{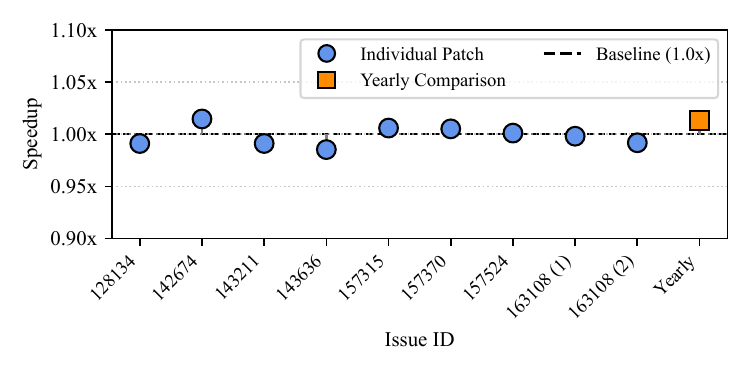}
\else%
    \includegraphics[width=\columnwidth]{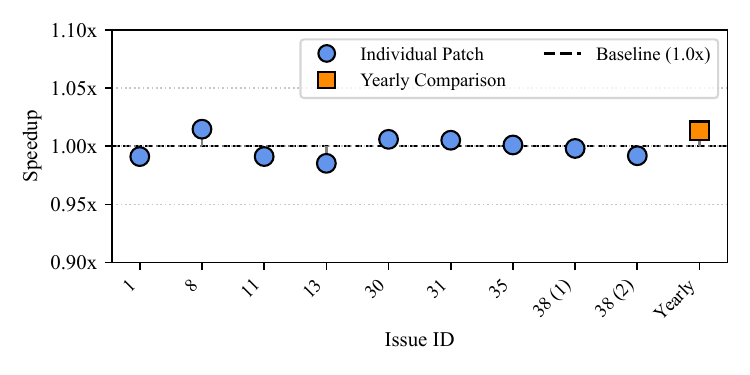}
\fi
    \caption{
        Geometric mean of speedup on SPEC CPU2017 Integer for selected
        patches, as well as for two LLVM versions
        (f3501d7 on Dec.\ 1, 2025 vs.\ 017c75b on Dec.\ 1, 2024).
    }
    \label{fig:patch-performance-ratio}
\end{figure}
\myparagraph{Runtime Efficiency on SPEC CPU2017}
To further assess the practical impact of the fixed missed optimizations,
we evaluated their runtime effects on the SPEC CPU2017 Integer benchmark
suite\footnote{We only included C/C++ benchmarks.}.
We focused on patches that are most likely to affect these
benchmarks.\footnote{We did not aggregate all patches into a single comparison,
as they were committed over time. Cherry-picking them onto a single LLVM version
is infeasible due to frequent changes in the related codebase.}
All experiments were conducted on a system equipped with an AMD Ryzen 9 7950X 16-core processor and 128 GB of RAM.
\cref{fig:patch-performance-ratio} reports the geometric mean speedup.
Following the instructions of the benchmark suite, each benchmark was executed three times,
and the median runtime was used for comparison~\cite{specOverview2017}.
The results indicate that none of the evaluated patches yields a significant speedup;
all observed performance differences are within 2\%, a range where measurement noise can have a substantial impact.
We speculate that this limited impact is likely due to the maturity of \llvm.
Having been developed and refined for decades, \llvm already incorporates a highly optimized compilation pipeline,
making it extremely challenging to achieve substantial performance improvements through a small number of
peephole optimizations alone. To further investigate this hypothesis,
we compared a recent \llvm version with one released approximately one year earlier;
the results are also shown in \cref{fig:patch-performance-ratio}.
This comparison similarly reveals no significant runtime performance difference,
lending additional support to our speculation.

\subsection{The Key Insight and Future Work}
The key insight of this work is that,
given a reliable and efficient mechanism for rigorously verifying the correctness of optimizations,
we can leverage the creativity of \llms to discover or even directly perform optimizations
without being hindered by their inherent unreliability.
In this work, we demonstrate the feasibility of this idea by using LLMs to detect
missed peephole optimizations in \llvm \ir, despite its maturity.
Moreover, the fact that \proj is able to identify \reportedInTotal missed peephole optimizations
suggests that state-of-the-art LLMs already possess sufficient code understanding and reasoning capabilities
to contribute meaningfully to compiler optimization tasks.

Several promising directions remain for future work.
First, although this work focuses on \llvm \ir,
the underlying idea also applies to other \ir
(e.g., MLIR) and instruction set architectures
(e.g., x86, ARM, and RISC-V).
Adapting \proj to these additional targets is therefore an important next step.
Second, \proj currently targets only peephole optimizations;
future studies could investigate whether \llms can also help identify missed cases
in other classes of optimizations.
Third, while \proj is designed to search for missed peephole optimizations,
the subsequent steps—generalizing optimization patterns and implementing them in the compiler—are
still performed manually by maintainers.
Future research could explore how LLMs might also assist with these generalization and implementation phases.
Finally, rigorous verification tools are essential to the success of \proj,
as they provide strong correctness guarantees for LLM-generated outputs.
Accordingly, future work may also focus on developing translation validation tools
for additional targets or enhancing existing tools (e.g., \alive).

\section{Related Work}
\label{sec:related_work}

In this section, we discuss the studies related to this work.

\myparagraph{Superoptimization}
A superoptimizer takes a sequence of instructions as input and searches for a more
efficient, semantically equivalent or refined program. Massalin~\cite{massalin1987superoptimizer}
first introduced the concept of superoptimization and developed the earliest
superoptimizer based on exhaustive enumeration, which, while pioneering, was both
inefficient and unsound. To address scalability limitations, subsequent work
proposed alternative search strategies, often leveraging SMT solvers for equivalence
verification. Joshi \etal developed Denali~\cite{joshi2002denali}, a goal-directed
superoptimizer that employs expert knowledge to encode equality-preserving
transformations, thereby constraining the search space. Bansal and
Aiken~\cite{bansal2006automatic} presented a superoptimizer that enumerates only
canonical instruction sequences to reduce redundancy and stores computed
optimizations in a database to avoid repeated work. Schkufza \etal introduced
STOKE~\cite{schkufza2013stochastic}, a superoptimizer for x86-64 binaries that
formulates superoptimization as a stochastic search problem and applies Markov Chain
Monte Carlo (MCMC) sampling to efficiently explore the vast program space. This
approach sacrifices completeness but significantly increases the diversity of
candidate programs, thereby improving the quality of the resulting code. Sasnauskas
\etal proposed \souper~\cite{sasnauskas2017souper}, a superoptimizer for \llvm IR that
extracts integer-typed functions from a module, traverses backward along dataflow
edges from the return instruction, and synthesizes more performant equivalents.
Souper is designed to be sound, but its applicability is limited to a purely
functional, control-flow-free subset of LLVM IR.
Liu \etal proposed
Minotaur~\cite{liu2024minotaur}, another
superoptimizer for \llvm IR that focuses on optimizations involving integer
and floating-point SIMD code. Minotaur supports more operations compared to \souper,
but its effectiveness is still constrained by the synthesis-based search strategy.
Hydra~\cite{mukherjee2024hydra} addresses the limitations of ungeneralized peephole
optimizations produced by superoptimizers by generalizing them using program synthesis.

\noindent\underline{Comparison with \proj: } Superoptimizers are capable of discovering new optimization patterns, but they are
computationally expensive, thus often limited to tiny window size or specific instruction sets.
They may also fail to produce results within reasonable time.
In contrast, \proj leverages \llms to explore optimization opportunities,
which is less restricted by the code size and instruction sets,
and thus can potentially detect more diverse missed optimizations.

\myparagraph{Differential Testing for Missed Optimizations}
Differential testing~\cite{mckeeman1998differential} compares the outputs of two or
more programs on identical inputs to detect discrepancies, which may reveal bugs.
Barany~\cite{barany2018finding} applied this technique to
compilers, identifying missed optimizations by comparing code generated by different
compilers for the same source program. Theodoridis \etal~\cite{theodoridis2022finding}
extended this approach to dead code elimination, while Liu \etal~\cite{liu2023exploring}
used differential testing to evaluate WebAssembly optimizers by comparing their
outputs with those of modern C compilers. Italiano and Cummins~\cite{italiano2024finding}
further advanced differential testing for code size optimizations, proposing methods
such as comparing outputs from seed and mutated code, evaluating different optimization
settings, and analyzing outputs from different compiler versions or distinct compilers.

\noindent\underline{Comparison with \proj: }
Differential testing is effective for discovering missed optimizations, but it
is limited to optimizations that have been implemented in other compilers
or that are not generalized enough.
In contrast, \proj can potentially discover new optimizations
that have not been implemented in any compiler.

\myparagraph{\llm-Based Approaches for Optimization}
Recent advances in \llms have
demonstrated their promise in program synthesis,
including code generation and optimization tasks.
A number of studies have explored the use of these models
for optimizations~\cite{garg2023rapgen, gao2024search, cummins2023llm, grubisic2024compiler, italiano2024finding}.
Some of them focus on source-level code optimization
and do not address the problem of finding missed optimizations~\cite{garg2023rapgen, gao2024search}.
Cummins \etal~\cite{cummins2023llm} investigated
the use of \llms to generate optimized LLVM IR and recommend compiler passes,
highlighting their potential but also their limitations
in terms of reusability and computational cost.
Grubisic \etal~\cite{grubisic2024compiler} incorporated feedback from optimizer-generated
code to enhance optimization quality.

\noindent\underline{Comparison with \proj: }
The only prior work that aims to detect missed optimizations with \llms
is by Italiano and Cummins~\cite{italiano2024finding}.
Their approach employs \llms to mutate code for differential testing,
achieving effective results but restricting the discovery of novel optimization patterns.
In contrast, \proj leverages the code understanding and reasoning capabilities of
\llms to directly detect suboptimal code snippets, which can potentially discover
new unimplemented optimization patterns.

\section{Conclusion}
\label{sec:conclusion}
This paper introduced \proj, a novel framework that bridges the creative
capabilities of \llm with the
rigorous guarantees of formal verification
to tackle the difficult challenge of discovering new peephole optimizations.
Our methodology overcomes the limitations of previous approaches by
integrating an \llm-based optimizer within a closed-loop system, where
verification feedback iteratively refines the \llm's search for correct and
efficient code transformations. Through a comprehensive evaluation on the
\llvm compiler, we demonstrated \proj's effectiveness in finding a substantial
number of previously unreported optimizations. The fact that many of these
have been confirmed and fixed in the official \llvm codebase highlights our
framework's practical significance and its ability to outperform
state-of-the-art superoptimizers.

Looking ahead, this research opens up exciting new avenues for the use of
\llms in compiler design and beyond. The success of \proj suggests that
similar hybrid methodologies, which combine the generative power of \llms
with rigorous formal methods, could be applied to other complex,
pattern-based optimization problems. As the reasoning abilities of \llms
continue to advance, we believe that frameworks like \proj will become an
indispensable tool for continuously improving the performance of modern
compilers, paving the way for more efficient and robust software.

\begin{acks}
We thank all the anonymous reviewers of ASPLOS'26 for their helpful
feedback and constructive comments.
We also thank the LLVM maintainers and contributors, especially Yingwei Zheng, for their efforts in
addressing the reported missed optimizations and for their valuable feedback.
This research is partially supported by
the Natural Sciences and Engineering Research Council of Canada
(NSERC) through the
Discovery Grant,
CFI-JELF Project \#40736,
and an unrestricted gift from Google.
\end{acks}

\bibliographystyle{ACM-Reference-Format}
\balance
\bibliography{acmart}

@inproceedings{theodoridis2022finding,
  title     = {Finding missed optimizations through the lens of dead code elimination},
  author    = {Theodoridis, Theodoros and Rigger, Manuel and Su, Zhendong},
  booktitle = {Proceedings of the 27th ACM International Conference on Architectural Support for Programming Languages and Operating Systems},
  pages     = {697--709},
  year      = {2022}
}

@inproceedings{liu2023exploring,
  title     = {Exploring missed optimizations in webassembly optimizers},
  author    = {Liu, Zhibo and Xiao, Dongwei and Li, Zongjie and Wang, Shuai and Meng, Wei},
  booktitle = {Proceedings of the 32nd ACM SIGSOFT International Symposium on Software Testing and Analysis},
  pages     = {436--448},
  year      = {2023}
}

@inproceedings{barany2018finding,
  title     = {Finding missed compiler optimizations by differential testing},
  author    = {Barany, Gerg{\"o}},
  booktitle = {Proceedings of the 27th international conference on compiler construction},
  pages     = {82--92},
  year      = {2018}
}

@article{italiano2024finding,
  title   = {Finding Missed Code Size Optimizations in Compilers using LLMs},
  author  = {Italiano, Davide and Cummins, Chris},
  journal = {arXiv preprint arXiv:2501.00655},
  year    = {2024}
}

@article{sasnauskas2017souper,
  title   = {Souper: A synthesizing superoptimizer},
  author  = {Sasnauskas, Raimondas and Chen, Yang and Collingbourne, Peter and Ketema, Jeroen and Lup, Gratian and Taneja, Jubi and Regehr, John},
  journal = {arXiv preprint arXiv:1711.04422},
  year    = {2017}
}

@article{schkufza2013stochastic,
  title     = {Stochastic superoptimization},
  author    = {Schkufza, Eric and Sharma, Rahul and Aiken, Alex},
  journal   = {ACM SIGARCH Computer Architecture News},
  volume    = {41},
  number    = {1},
  pages     = {305--316},
  year      = {2013},
  publisher = {ACM New York, NY, USA}
}

@article{massalin1987superoptimizer,
  title        = {Superoptimizer: a look at the smallest program},
  volume       = {15},
  rights       = {https://www.acm.org/publications/policies/copyright_policy#Background},
  issn         = {0163-5964},
  doi          = {10.1145/36177.36194},
  abstractnote = {Given an instruction set, the superoptimizer finds the shortest program to compute a function. Startling programs have been generated, many of them engaging in convoluted bit-fiddling bearing little resemblance to the source programs which defined the functions. The key idea in the superoptimizer is a probabilistic test that makes exhaustive searches practical for programs of useful size. The search space is defined by the processor’s instruction set, which may include the whole set, but it is typically restricted to a subset. By constraining the instructions and observing the effect on the output program, one can gain insight into the design of instruction sets. In addition, superoptimized programs may be used by peephole optimizers to improve the quality of generated code, or by assembly language programmers to improve manually written code.},
  number       = {5},
  journal      = {ACM SIGARCH Computer Architecture News},
  publisher    = {Association for Computing Machinery (ACM)},
  author       = {Massalin, Henry},
  year         = {1987},
  month        = nov,
  pages        = {122–126},
  language     = {en}
}

@inproceedings{davidson1984automatic,
  address   = {Montreal, Canada},
  title     = {Automatic generation of peephole optimizations},
  rights    = {https://www.acm.org/publications/policies/copyright_policy#Background},
  url       = {http://portal.acm.org/citation.cfm?doid=502874.502885},
  doi       = {10.1145/502874.502885},
  booktitle = {Proceedings of the 1984 SIGPLAN symposium on Compiler construction  - SIGPLAN ’84},
  publisher = {ACM Press},
  author    = {Davidson, Jack W. and Fraser, Christopher W.},
  year      = {1984},
  pages     = {111–116}
}

@article{mckeeman1965peephole,
  title        = {Peephole optimization},
  volume       = {8},
  rights       = {https://www.acm.org/publications/policies/copyright_policy#Background},
  issn         = {0001-0782, 1557-7317},
  doi          = {10.1145/364995.365000},
  abstractnote = {Redundant instructions may be discarded during the final stage of compilation by using a simple optimizing technique called peephole optimization. The method is described and examples are given.},
  number       = {7},
  journal      = {Communications of the ACM},
  publisher    = {Association for Computing Machinery (ACM)},
  author       = {McKeeman, W. M.},
  year         = {1965},
  month        = july,
  pages        = {443–444},
  language     = {en}
}

@inproceedings{bansal2006automatic,
  address   = {San Jose California USA},
  title     = {Automatic generation of peephole superoptimizers},
  rights    = {https://www.acm.org/publications/policies/copyright_policy#Background},
  url       = {https://dl.acm.org/doi/10.1145/1168857.1168906},
  doi       = {10.1145/1168857.1168906},
  booktitle = {Proceedings of the 12th international conference on Architectural support for programming languages and operating systems},
  publisher = {ACM},
  author    = {Bansal, Sorav and Aiken, Alex},
  year      = {2006},
  month     = oct,
  pages     = {394–403}
}

@article{joshi2002denali,
  title        = {Denali: a goal-directed superoptimizer},
  volume       = {37},
  rights       = {https://www.acm.org/publications/policies/copyright_policy#Background},
  issn         = {0362-1340, 1558-1160},
  doi          = {10.1145/543552.512566},
  abstractnote = {This paper provides a preliminary report on a new research project that aims to construct a code generator that uses an automatic theorem prover to produce very high-quality (in fact, nearly mathematically optimal) machine code for modern architectures. The code generator is not intended for use in an ordinary compiler, but is intended to be used for inner loops and critical subroutines in those cases where peak performance is required, no available compiler generates adequately efficient code, and where current engineering practice is to use hand-coded machine language. The paper describes the design of the superoptimizer, and presents some encouraging preliminary results.},
  number       = {5},
  journal      = {ACM SIGPLAN Notices},
  publisher    = {Association for Computing Machinery (ACM)},
  author       = {Joshi, Rajeev and Nelson, Greg and Randall, Keith},
  year         = {2002},
  month        = may,
  pages        = {304–314},
  language     = {en}
}

@article{mukherjee2024hydra,
  title        = {Hydra: Generalizing Peephole Optimizations with Program Synthesis},
  volume       = {8},
  rights       = {https://creativecommons.org/licenses/by/4.0/},
  issn         = {2475-1421},
  doi          = {10.1145/3649837},
  abstractnote = {Optimizing compilers rely on peephole optimizations to simplify  combinations of instructions and remove redundant instructions.  Typically, a new peephole optimization is added when a compiler  developer notices an optimization opportunity---a collection of  dependent instructions that can be improved---and manually derives a  more general rewrite rule that optimizes not only the original code,  but also other, similar collections of instructions.  In this paper, we present Hydra, a tool that automates the process of  generalizing peephole optimizations using a collection of techniques  centered on program synthesis.  One of the most important problems we have solved is finding a version  of each optimization that is independent of the bitwidths of the  optimization’s inputs (when this version exists).  We show that Hydra can generalize 75% of the ungeneralized missed  peephole optimizations that LLVM developers have posted to the LLVM  project’s issue tracker.  All of Hydra’s generalized peephole optimizations have been formally  verified, and furthermore we can automatically turn them into C++ code  that is suitable for inclusion in an LLVM pass.},
  number       = {OOPSLA1},
  journal      = {Proceedings of the ACM on Programming Languages},
  publisher    = {Association for Computing Machinery (ACM)},
  author       = {Mukherjee, Manasij and Regehr, John},
  year         = {2024},
  month        = apr,
  pages        = {725–753},
  language     = {en}
}

@article{mckeeman1998differential,
  title   = {Differential testing for software},
  author  = {McKeeman, William M},
  journal = {Digital Technical Journal},
  volume  = {10},
  number  = {1},
  pages   = {100--107},
  year    = {1998}
}

@article{cummins2023llm,
  title        = {Large Language Models for Compiler Optimization},
  url          = {http://arxiv.org/abs/2309.07062},
  doi          = {10.48550/arXiv.2309.07062},
  abstractnote = {We explore the novel application of Large Language Models to code optimization. We present a 7B-parameter transformer model trained from scratch to optimize LLVM assembly for code size. The model takes as input unoptimized assembly and outputs a list of compiler options to best optimize the program. Crucially, during training, we ask the model to predict the instruction counts before and after optimization, and the optimized code itself. These auxiliary learning tasks significantly improve the optimization performance of the model and improve the model’s depth of understanding. We evaluate on a large suite of test programs. Our approach achieves a 3.0% improvement in reducing instruction counts over the compiler, outperforming two state-of-the-art baselines that require thousands of compilations. Furthermore, the model shows surprisingly strong code reasoning abilities, generating compilable code 91% of the time and perfectly emulating the output of the compiler 70% of the time.},
  note         = {arXiv:2309.07062 [cs]},
  number       = {arXiv:2309.07062},
  publisher    = {arXiv},
  author       = {Cummins, Chris and Seeker, Volker and Grubisic, Dejan and Elhoushi, Mostafa and Liang, Youwei and Roziere, Baptiste and Gehring, Jonas and Gloeckle, Fabian and Hazelwood, Kim and Synnaeve, Gabriel and Leather, Hugh},
  year         = {2023},
  month        = sep
}

@inbook{buchwald2015optgen,
  address      = {Berlin, Heidelberg},
  series       = {Lecture Notes in Computer Science},
  title        = {Optgen: A Generator for Local Optimizations},
  volume       = {9031},
  rights       = {http://www.springer.com/tdm},
  isbn         = {978-3-662-46662-9},
  url          = {http://link.springer.com/10.1007/978-3-662-46663-6_9},
  doi          = {10.1007/978-3-662-46663-6_9},
  abstractnote = {Every compiler comes with a set of local optimization rules, such as x + 0 → x and x & x → x, that do not require any global analysis. These rules reﬂect the wisdom of the compiler developers about mathematical identities that hold for the operations of their intermediate representation. Unfortunately, these sets of hand-crafted rules guarantee neither correctness nor completeness. Optgen solves this problem by generating all local optimizations up to a given cost limit. Since Optgen veriﬁes each rule using an SMT solver, it guarantees correctness and completeness of the generated rule set. Using Optgen, we tested the latest versions of GCC, ICC and LLVM and identiﬁed more than 50 missing local optimizations that involve only two operations.},
  booktitle    = {Compiler Construction},
  publisher    = {Springer Berlin Heidelberg},
  author       = {Buchwald, Sebastian},
  editor       = {Franke, Björn},
  year         = {2015},
  pages        = {171–189},
  collection   = {Lecture Notes in Computer Science},
  language     = {en}
}

@book{fischer1991crafting,
  title     = {Crafting a Compiler with C},
  author    = {Fischer, Charles N and LeBlanc Jr, Richard J},
  year      = {1991},
  publisher = {Benjamin-Cummings Publishing Co., Inc.}
}

@phdthesis{lattner2002llvm,
  title  = {LLVM: An infrastructure for multi-stage optimization},
  author = {Lattner, Chris Arthur},
  year   = {2002},
  school = {University of Illinois at Urbana-Champaign}
}

@inproceedings{lattner2004llvm,
  title        = {LLVM: A compilation framework for lifelong program analysis \& transformation},
  author       = {Lattner, Chris and Adve, Vikram},
  booktitle    = {International symposium on code generation and optimization, 2004. CGO 2004.},
  pages        = {75--86},
  year         = {2004},
  organization = {IEEE}
}

@inproceedings{lopes2015provably,
  title     = {Provably correct peephole optimizations with alive},
  author    = {Lopes, Nuno P and Menendez, David and Nagarakatte, Santosh and Regehr, John},
  booktitle = {Proceedings of the 36th ACM SIGPLAN Conference on Programming Language Design and Implementation},
  pages     = {22--32},
  year      = {2015}
}

@inproceedings{lopes2021alive2,
  title     = {Alive2: bounded translation validation for LLVM},
  author    = {Lopes, Nuno P and Lee, Juneyoung and Hur, Chung-Kil and Liu, Zhengyang and Regehr, John},
  booktitle = {Proceedings of the 42nd ACM SIGPLAN International Conference on Programming Language Design and Implementation},
  pages     = {65--79},
  year      = {2021}
}

@inproceedings{pnueli1998translation,
  title        = {Translation validation},
  author       = {Pnueli, Amir and Siegel, Michael and Singerman, Eli},
  booktitle    = {International Conference on Tools and Algorithms for the Construction and Analysis of Systems},
  pages        = {151--166},
  year         = {1998},
  organization = {Springer}
}

@inproceedings{necula2000translation,
  title     = {Translation validation for an optimizing compiler},
  author    = {Necula, George C},
  booktitle = {Proceedings of the ACM SIGPLAN 2000 conference on Programming language design and implementation},
  pages     = {83--94},
  year      = {2000}
}

@article{grubisic2024compiler,
  title   = {Compiler generated feedback for large language models},
  author  = {Grubisic, Dejan and Cummins, Chris and Seeker, Volker and Leather, Hugh},
  journal = {arXiv preprint arXiv:2403.14714},
  year    = {2024}
}

@article{garg2023rapgen,
  title   = {Rapgen: An approach for fixing code inefficiencies in zero-shot},
  author  = {Garg, Spandan and Moghaddam, Roshanak Zilouchian and Sundaresan, Neel},
  journal = {arXiv preprint arXiv:2306.17077},
  year    = {2023}
}

@article{gao2024search,
  title   = {Search-based llms for code optimization},
  author  = {Gao, Shuzheng and Gao, Cuiyun and Gu, Wenchao and Lyu, Michael},
  journal = {arXiv preprint arXiv:2408.12159},
  year    = {2024}
}

@inproceedings{DBLP:conf/osdi/BansalA08,
  author    = {Sorav Bansal and
               Alex Aiken},
  editor    = {Richard Draves and
               Robbert van Renesse},
  title     = {Binary Translation Using Peephole Superoptimizers},
  booktitle = {8th {USENIX} Symposium on Operating Systems Design and Implementation,
               {OSDI} 2008, December 8-10, 2008, San Diego, California, USA, Proceedings},
  pages     = {177--192},
  publisher = {{USENIX} Association},
  year      = {2008},
  url       = {http://www.usenix.org/events/osdi08/tech/full\_papers/bansal/bansal.pdf},
  bibsource = {dblp computer science bibliography, https://dblp.org}
}

@book{dragonBook,
  author    = {Aho, Alfred V. and Lam, Monica S. and Sethi, Ravi and Ullman, Jeffrey D.},
  isbn      = {0321486811},
  month     = {August},
  publisher = {{Addison Wesley}},
  title     = {Compilers: Principles, Techniques, and Tools (2nd Edition)},
  year      = 2006
}

@misc{llvm2025llvm,
  author       = {},
  title        = {{T}he {L}{L}{V}{M} {C}ompiler {I}nfrastructure {P}roject --- llvm.org},
  howpublished = {\url{https://llvm.org}},
  year         = {[n.\,d.]},
  note         = {Accessed 01-08-2025}
}

@misc{llvm2025languageref,
  author       = {},
  title        = {{L}{L}{V}{M} {L}anguage {R}eference {M}anual --- {L}{L}{V}{M} 22.0.0git documentation --- llvm.org},
  howpublished = {\url{https://llvm.org/docs/LangRef.html}},
  year         = {[n.\,d.]},
  note         = {Accessed 01-08-2025}
}

@misc{llvm2025instcombine,
  author       = {},
  title        = {{I}nst{C}ombine contributor guide --- {L}{L}{V}{M} 22.0.0git documentation --- llvm.org},
  howpublished = {\url{https://llvm.org/docs/InstCombineContributorGuide.html}},
  year         = {[n.\,d.]},
  note         = {Accessed 01-08-2025}
}

@misc{instcombinepr,
  author       = {},
  title        = {LLVM InstCombine Pull Requests on GitHub},
  howpublished = {\url{https://github.com/llvm/llvm-project/pulls?q=is\%3Aopen+is\%3Apr+label\%3Allvm\%3Ainstcombine}},
  year         = {[n.\,d.]},
  note         = {Accessed 01-08-2025}
}

@article{DBLP:journals/corr/abs-2502-06215,
  author     = {Xin Zhou and
                Martin Weyssow and
                Ratnadira Widyasari and
                Ting Zhang and
                Junda He and
                Yunbo Lyu and
                Jianming Chang and
                Beiqi Zhang and
                Dan Huang and
                David Lo},
  title      = {LessLeak-Bench: {A} First Investigation of Data Leakage in LLMs Across
                83 Software Engineering Benchmarks},
  journal    = {CoRR},
  volume     = {abs/2502.06215},
  year       = {2025},
  url        = {https://doi.org/10.48550/arXiv.2502.06215},
  doi        = {10.48550/ARXIV.2502.06215},
  eprinttype = {arXiv},
  eprint     = {2502.06215},
  bibsource  = {dblp computer science bibliography, https://dblp.org}
}

@article{DBLP:journals/tse/LopezCSSV25,
  author    = {Jos{\'{e}} Antonio Hern{\'{a}}ndez L{\'{o}}pez and
               Boqi Chen and
               Mootez Saad and
               Tushar Sharma and
               D{\'{a}}niel Varr{\'{o}}},
  title     = {On Inter-Dataset Code Duplication and Data Leakage in Large Language
               Models},
  journal   = {{IEEE} Trans. Software Eng.},
  volume    = {51},
  number    = {1},
  pages     = {192--205},
  year      = {2025},
  url       = {https://doi.org/10.1109/TSE.2024.3504286},
  doi       = {10.1109/TSE.2024.3504286},
  bibsource = {dblp computer science bibliography, https://dblp.org}
}

@inproceedings{menendez2016termination,
  title     = {Termination-checking for LLVM peephole optimizations},
  author    = {Menendez, David and Nagarakatte, Santosh},
  booktitle = {Proceedings of the 38th International Conference on Software Engineering},
  pages     = {191--202},
  year      = {2016}
}

@misc{llvm-opt-benchmark,
  title        = {{LLVM Opt Benchmark}},
  howpublished = {\url{https://github.com/dtcxzyw/llvm-opt-benchmark}},
  author       = {Yingwei Zheng},
  year         = {2023},
  note         = {Accessed: 30-12-2025}
}

@article{liu2024minotaur,
  title     = {Minotaur: A SIMD-oriented synthesizing superoptimizer},
  author    = {Liu, Zhengyang and Mada, Stefan and Regehr, John},
  journal   = {Proceedings of the ACM on Programming Languages},
  volume    = {8},
  number    = {OOPSLA2},
  pages     = {1561--1585},
  year      = {2024},
  publisher = {ACM New York, NY, USA}
}

@misc{specOverview2017,
  author       = {},
  title        = {{O}verview - {C}{P}{U} 2017 --- spec.org},
  howpublished = {\url{https://www.spec.org/cpu2017/Docs/overview.html}},
  year         = {[n.\,d.]},
  note         = {Accessed 12-01-2026}
}

@misc{llvm-compile-time-tracker,
  author = {Nikita Popov},
  title  = {llvm-compile-time-tracker: LLVM compile-time performance tracking infrastructure},
  year   = {2025},
  url    = {https://github.com/nikic/llvm-compile-time-tracker}
}

\end{document}